%% file: aholom.tex
\documentclass[letterpaper,10pt,twoside]{article}
\usepackage[utf8]{inputenc}
\usepackage[english]{babel}
\usepackage[T1]{fontenc}
\usepackage{lmodern}
\usepackage{amsmath,amssymb,amsthm} 
\usepackage[ps,arrow,matrix]{xy} 

\newcommand{\im}{\mathrm{i}}
\newcommand{\N}{\mathbb{N}}

\newcommand{\R}{\mathbb{R}}
\newcommand{\C}{\mathbb{C}}

\newcommand{\defeq}{:=}
\newcommand{\tens}{\otimes}
\newcommand{\ctens}{\hat{\otimes}}
\newcommand{\cH}{\mathcal{H}}
\newcommand{\rH}{\mathrm{H}}
\newcommand{\cL}{\mathcal{L}}
\newcommand{\xd}{\mathrm{d}}

\newcommand{\cont}{\mathrm{C}}
\newcommand{\toi}{\hookrightarrow}

\newcommand{\id}{\mathrm{id}}

\newcommand{\ds}{\circ}
\newcommand{\ano}[1]{\blacktriangleleft #1 \blacktriangleright}

\theoremstyle{definition}
\newtheorem{dfn}{Definition}[section]

\theoremstyle{plain}
\newtheorem{lem}[dfn]{Lemma}
\newtheorem{prop}[dfn]{Proposition}
\newtheorem{thm}[dfn]{Theorem}

\allowdisplaybreaks

\begin{document}

\input{titlepage}

\input{intro}

\input{ingreds}

\input{gaxioms}

\input{classdata}

\input{caxioms}

\input{quantization}

\input{prop}

\input{linsrc}

\input{outlook}

\bibliographystyle{amsordx} 
\bibliography{stdrefs}
\end{document}

%% file: titlepage.tex
\begin{titlepage}
\title{\textbf{Affine holomorphic quantization}}
\author{Robert Oeckl\footnote{email: robert@matmor.unam.mx}\\ \\
Instituto de Matemáticas,\\
Universidad Nacional Autónoma de México,\\
Campus Morelia, C.P.~58190, Morelia, Michoacán, Mexico}
\date{UNAM-IM-MOR-2011-1\\ 28 April 2011\\ 16 February 2012 (v2)\\ 7 September 2012 (v3)}

\maketitle

\vspace{\stretch{1}}

\begin{abstract}
\input{abstract}
\end{abstract}

\vspace{\stretch{1}}
\end{titlepage}

%% file: abstract.tex
We present a rigorous and functorial quantization scheme for affine field theories, i.e., field theories where local spaces of solutions are affine spaces. The target framework for the quantization is the general boundary formulation, allowing to implement manifest locality without the necessity for metric or causal background structures. The quantization combines the holomorphic version of geometric quantization for state spaces with the Feynman path integral quantization for amplitudes. We also develop an adapted notion of coherent states, discuss vacuum states, and consider observables and their Berezin-Toeplitz quantization. Moreover, we derive a factorization identity for the amplitude in the special case of a linear field theory modified by a source-like term and comment on its use as a generating functional for a generalized S-matrix.

%% file: intro.tex
\section{Introduction}

Ever since its inception, efforts have been made to put quantum field theory on an axiomatic basis. There are multiple objectives behind such undertakings. Conceptually, one would like to have a better understanding of what quantum field theory ``really is'' (and what it is not), possibly including an elucidation of aspects of the meaning or interpretation of quantum theory itself. Mathematically, an axiomatic system offers a rigorous definition and a context to make mathematically precise statements about certain quantum field theories or quantum field theory as such. Finally, an axiomatic formulation may help to indicate how quantum field theories can be extended to realms where they have not previously been experimentally tested. An important example for the latter is the extension from Minkowski space to more general curved spacetime.

An axiomatic approach that has proven particularly useful in this latter respect
is \emph{algebraic quantum field theory} (AQFT) \cite{Haa:lqp}.
In AQFT the causal structure of spacetime is intimately entwined with the algebraic structure of the objects of the quantum theory.
This has advantages and disadvantages. Most notably, this leads to a very concise way of encoding local physics in a spacetime region, with just one core mathematical structure (a von~Neumann or $C^*$ algebra) per spacetime region. Moreover, in quantization prescriptions this structure is directly linked to the classical observables in that spacetime region. This conciseness combined with mathematical rigor has justifiably fascinated physicists and mathematicians over the decades, making it today the best developed axiomatic approach to quantum field theory.

On the other hand, the central role played by causality in the core structure of AQFT makes it indispensable as a fixed ingredient of spacetime. This precludes the direct applicability of AQFT to situations where such a structure is not a priori given.

This limitation, which is even more stringent in most other approaches to quantum field theory, has motivated a new axiomatic approach, called the \emph{general boundary formulation} (GBF). The GBF has been put forward with the express aim of disentangling the elementary mathematical objects of a theory (in this case states, amplitudes, observables) and their basic physical interpretation, from the metric or causal structure of spacetime.
This is achieved on the one hand by explicitly localizing states on hypersurfaces and amplitudes in spacetime regions \cite{Oe:boundary} in the spirit of topological quantum field theory \cite{Ati:tqft}. On the other hand this requires an extension of the probability postulates of quantum theory for amplitudes \cite{Oe:gbqft} and observables \cite{Oe:obsgbf}.
While still considerably less developed than, say, AQFT, the GBF offers the perspective of further extending the realm of quantum field theory to contexts where spacetime is not equipped with a predetermined metric or causal background structure. It is widely expected that a quantum theory of gravity should live precisely in such a ``background independent'' context.

Most realistic quantum field theories are obtained or at least motivated through a process of quantization starting with a classical field theory. It is thus important for the usefulness of a given axiomatic approach that there be quantization prescriptions that produce the elementary objects which are the subject of the axioms starting from data encoding a classical field theory. In the case of the GBF the quantization prescription most straightforwardly adapted from well known tools of quantum (field) theory is Schrödinger-Feynman quantization \cite{Oe:gbqft,Oe:kgtl}, which combines the Schrödinger representation \cite{Jac:schroedinger} for state spaces with the Feynman path integral \cite{Fey:stnrqm} for amplitudes. This quantization prescription has been successfully applied in various contexts including a non-perturbative integrable model \cite{Oe:2dqym}, a generalization of the perturbative S-matrix \cite{CoOe:smatrixgbf}, and in curved spacetime \cite{CoOe:unitary,Col:desitterpaper}. Even though many of these applications lead to structures that rigorously satisfy the axioms, the quantization prescription itself is not rigorously formulated, at least not in its present form.

Ideally, quantization should not only be rigorous, but should provide something like a functor from a category of classical theories to a category of quantum theories. For the GBF such a functorial quantization scheme has indeed been described recently for the case of linear field theory \cite{Oe:holomorphic}. There, the concept of a linear classical field theory is axiomatized and a construction is given that produces from the elementary objects of such a classical theory the elementary objects of a quantum field theory in the framework of the GBF. In particular, it is proven that the objects of the quantum theory obtained in this way do indeed satisfy the axioms of the GBF. Moreover, although it is not made explicit there, this construction is functorial, and in many ways so. For example, for a given system of spacetime hypersurfaces and regions we obtain a functor if we take the categories of classical and quantum field theories with morphisms given by the respective notion of ``subtheory'': On the classical side a ``subtheory'' is obtained by restricting the local spaces of solutions consistently to subspaces, while on the quantum side a ``subtheory'' is obtained by decomposing the local Hilbert spaces of states into tensor products and selecting one component in a consistent way. Other possibilities for choices of categories include ones where each object carries its own system of hypersurfaces and regions etc.

A classical linear field theory is formalized in \cite{Oe:holomorphic} as follows: For each region in spacetime we are given a real vector space of solutions of the field equations. Also, for each hypersurface in spacetime we are given a real vector space of germs of solutions. The latter spaces are moreover equipped with non-degenerate symplectic forms. Then, the natural maps from the former spaces to the latter (restricting solutions in regions to neighborhoods of the boundary) have to yield Lagrangian subspaces with respect to these symplectic forms. Although perhaps not obviously so, these conditions are well motivated from Lagrangian field theory. An additional ingredient which might be seen as structure already pertaining to the quantum realm is a compatible complex structure on the solution space for each hypersurface. This summarizes the axioms given in \cite{Oe:holomorphic} for a classical linear field theory in an informal language.

The quantization prescription consists then of a combination of a version of geometric quantization for hypersurfaces and a certain integral quantization for regions. For each hypersurface, the construction of the associated Hilbert space of states is equivalent to the usual Fock space construction, where the phase space (here really the space of germs of classical solutions in a neighborhood of the hypersurface) with additional symplectic and complex structure is seen as the (dual of the) 1-particle Hilbert space. However, it is realized concretely as a space of holomorphic functions in the spirit of Bargmann. From the point of view of geometric quantization this is really the space of Kähler polarized sections of the prequantum bundle. For each region, the quantization prescription in \cite{Oe:holomorphic} is given by a seemingly ad~hoc integral prescription, although verified by providing the ``right'' results in certain examples.

In the present paper we consider \emph{affine field theory}, as a first case of a rigorous and functorial quantization prescription targeting the GBF beyond linear field theory. By affine field theory we mean here field theory with affine spaces of local solutions and such that the natural symplectic forms associated to hypersurfaces are invariant with respect to the affine structure in addition to being non-degenerate. In many ways this can be seen as a generalization of the linear case and its treatment in \cite{Oe:holomorphic}. For hypersurfaces, this requires a refinement of the geometric quantization prescription (Section~\ref{sec:geomquant}), clarifying the role of the prequantum bundle and its relevant trivializations. For regions, we motivate the quantization as a variant of the Feynman path integral prescription (Section~\ref{sec:feynquant}), thus justifying at the same time the origin of the prescription given in \cite{Oe:holomorphic} as a special case of this.

Based on a suitable geometric setting for spacetime (Section~\ref{sec:geomax}), the axioms for classical field theory (Section~\ref{sec:classdata}) are a relatively straightforward generalization of those for linear field theory given in \cite{Oe:holomorphic}. However, they involve additional structural elements from Lagrangian field theory (see Sections~\ref{sec:classft} and \ref{sec:affine}), notably the action and the symplectic potential. Also, they are considerably more extensive as both local spaces of solutions and their tangent spaces need to be kept track of separately since they are no longer canonically identified.

The central part of this paper is Section~\ref{sec:quantization} where the quantization prescription is specified rigorously and the validity of the GBF core axioms (listed in Section~\ref{sec:coreaxioms}) is proven. As in \cite{Oe:holomorphic} the Hilbert spaces of states associated to hypersurfaces are realized concretely as spaces of functions (Section~\ref{sec:sspaces}). However, the domain spaces (or rather their extensions) for these functions do not directly carry measures as in \cite{Oe:holomorphic}. Rather, any choice of base point gives rise to an identification with a space of holomorphic functions with a measure on (an extension of) the domain space. This is then used to obtain the inner product, which turns out to be independent of the base point. In terms of geometric quantization these different function spaces arise from different trivializations of the prequantum bundle.

In Section~\ref{sec:cohstates}, coherent states are defined. These are called \emph{affine coherent states} to distinguish them from the usual coherent states (used in \cite{Oe:holomorphic}). While the latter can also be ``imported'' into the affine setting, their definition and properties are base point dependent and 
therefore less convenient than the manifestly base point independent affine coherent states. In Section~\ref{sec:ampl} the amplitude functions are defined and some of their elementary properties considered. In particular, an explicit formula for the amplitude of coherent states is obtained, generalizing the corresponding result from the case of linear field theory. Section~\ref{sec:gluing} provides a proof of the gluing axiom, with the other GBF core axioms already proven in the previous sections. As in \cite{Oe:holomorphic} the proof of this last axiom requires an additional integrability condition on the classical data.

In Section~\ref{sec:prop} some further aspects of the proposed quantization prescription are considered: Section~\ref{sec:evol} discusses some aspects of the picture that emerges if we choose to focus on amplitudes that may be viewed as ``transition'' amplitudes in a context of ``evolution'' between hypersurfaces. In Section~\ref{sec:vacuum} vacua in the sense of \cite{Oe:gbqft} are discussed. Unsurprisingly, there is no longer a preferred vacuum in the affine theory as there is in the linear theory. Nevertheless the finding in \cite{Oe:holomorphic} that each global solution of the classical theory gives rise to a vacuum remains true in the affine setting. The relation between the linear and the affine setting on the quantum level is clarified in Section~\ref{sec:redlin}. Observables in the sense of \cite{Oe:obsgbf} are discussed in Section~\ref{sec:observables}. In particular, the Berezin-Toeplitz quantization of observables given in \cite{Oe:obsgbf} for the linear setting is generalized to the affine setting, including a generalization of the coherent factorization property.

Finally, in Section~\ref{sec:linsrc} we consider in some detail a special case of particular interest. A linear field theory is given in a spacetime region. In the interior of that region a linear term is added to the action making the theory there affine. We are then interested in describing this affine theory in terms of the original linear theory. This turns out to lead to a remarkable factorization of the amplitude of the affine theory (Section~\ref{sec:factampl}). An important example for this setting is the case where the linear addition to the action is a source term (Section~\ref{sec:musrc}) in which case the resulting amplitude may be seen as leading to a generator of the perturbative S-matrix. In an evolution picture (Section~\ref{sec:evolal}) one recovers a generalization of the well known particle creation from the vacuum through a source.

Section~\ref{sec:outlook} presents a brief outlook.

Coming back to issues mentioned at the beginning of this section, we stress that all constructions and results of this paper (except where explicitly stated otherwise) apply to field theory understood in a rather abstract and general sense. In particular, nowhere do we need to assume a particular spacetime metric or causal structure or even the existence of such a structure.

Various constructions in Section~\ref{sec:quantization} as well as most proofs in this paper build on results of \cite{Oe:holomorphic}, to which we refer the interested reader for those details.

%% file: ingreds.tex
\section{Motivation of quantization scheme}
\label{sec:ingredients}

The quantization scheme put forward in this paper may be seen as a combination of a geometric quantization (for state spaces) with a Feynman path integral quantization (for amplitudes). We proceed to explain this in the present section.

\subsection{Ingredients from classical field theory}
\label{sec:classft}

We recall certain elementary ingredients of Lagrangian field theory here, relying on the conventions and notation in \cite{Oe:holomorphic}. Thus, we suppose a classical field theory to be defined on a smooth spacetime manifold $T$ of dimension $d$ and determined by a first order Lagrangian density $\Lambda(\varphi,\partial\varphi,x)$ with values in $d$-forms on $T$. Here $x\in T$ denotes a point in spacetime, $\varphi$ a field configuration at a point and $\partial\varphi$ the spacetime derivative at a point of a field configuration. We shall assume that the configurations are sections of a trivial vector bundle over $T$. We shall also assume in the following that all fields decay sufficiently rapidly at infinity where required (i.e., where regions or hypersurfaces are non-compact).

Given a spacetime region $M$ and a field configuration $\phi$ in $M$ its \emph{action} is given by
\begin{equation}
 S_M(\phi)\defeq \int_M \Lambda(\phi(\cdot),\partial\phi(\cdot),\cdot) .
\end{equation}
$S_M$ is usually viewed as a real valued function on the space of field configurations on $M$. However, in the following we will often be interested only in the value of $S_M$ on the space $A_M$ of solutions of the Euler-Lagrange equations in $M$.
Given a hypersurface $\Sigma$ we denote by $A_\Sigma$ the space of (germs of) solutions of the Euler-Lagrange equations in a neighborhood of $\Sigma$. The \emph{symplectic potential} is then the one-form on $A_\Sigma$ defined as
\begin{equation}
 (\theta_{\Sigma})_{\phi}(X)\defeq -\int_\Sigma X^a \left.\partial_\mu\lrcorner\frac{\delta \Lambda}{\delta\, \partial_\mu\varphi^a}\right|_\phi .
\label{eq:sympot}
\end{equation}
Here $\phi\in A_\Sigma$ while $X$ is a tangent vector to $\phi$, i.e., an element of the space $T_\phi A_\Sigma$ of solutions linearized around $\phi$.
The restriction of solutions in the interior of a region $M$ to its boundary $\partial M$ induces a map $a_M:A_M\to A_{\partial M}$. Given $\phi\in A_M$ this induces a map between linearized solutions $(a_{M}^*)_\phi:T_\phi A_M\to T_{a_M(\phi)} A_{\partial M}$. The symplectic potential is then related to the exterior derivative of the action via
\begin{equation}
 (\theta_{\partial M})_{a_M(\phi)}((a_M^*)_{\phi}(X))=-(\xd S_M)_\phi(X) .
\label{eq:relspact}
\end{equation}
For a hypersurface $\Sigma$, the \emph{symplectic form} is the two-form on $A_\Sigma$ given by the exterior derivative of the symplectic potential, 
\begin{multline}
(\omega_\Sigma)_\phi(X,Y)  =(\xd\theta_\Sigma)_\phi(X,Y)
 =-\frac{1}{2}\int_\Sigma\left( (X^b Y^a-Y^b X^a)\left.\partial_\mu\lrcorner
 \frac{\delta^2\Lambda}{\delta\varphi^b\delta\,\partial_\mu\varphi^a}\right|_\phi\right. \\
 \left. + (Y^a\partial_\nu X^b-X^a \partial_\nu Y^b)\left.\partial_\mu\lrcorner
 \frac{\delta^2\Lambda}{\delta\,\partial_\nu\varphi^b\delta\,\partial_\mu\varphi^a}\right|_\phi\right) .
\label{eq:sympl}
\end{multline}
We shall assume that the symplectic structure is always non-degenerate.

Note that a change of orientation of the hypersurface $\Sigma$ changes the sign of the symplectic potential in (\ref{eq:sympot}) and consequently that of the symplectic form in (\ref{eq:sympl}). In quantization schemes that consider a global space of solutions in $T$, the orientation of the (then usually spacelike) hypersurface $\Sigma$ has no particular importance and the sign of the symplectic potential and the symplectic form is chosen in a manner convenient for the quantization. Indeed, in text books the formulas (\ref{eq:sympot}) and (\ref{eq:sympl}) are often presented with the opposite sign. For our purposes, however, the choice of sign turns out to be uniquely determined by the interplay between geometric quantization on hypersurfaces and Feynman quantization on regions. The key is here the relative sign in equation (\ref{eq:relspact}). We will come back to this issue in Section~\ref{sec:ampl}.

Recall that given a region $M$ and a solution $\phi\in A_M$, the space $T_\phi A_M$ of solutions linearized around $\phi$ is an isotropic subspace of $T_{a_M(\phi)} A_{\partial M}$ as follows by taking the exterior derivative on both sides of (\ref{eq:relspact}) and noticing that $\xd\xd S_M=0$. In many cases of interest this subspace is also coisotropic and hence Lagrangian, see \cite{Oe:holomorphic} for further remarks on this point.

\subsection{Specializing to affine field theory}
\label{sec:affine}

We specialize now to the type of field theory of principal interest in the present paper: \emph{affine field theory}. We proceed to explain exactly what we mean by this. Firstly, we suppose that the spaces of solutions $A_M$ for regions $M$ and $A_\Sigma$ for hypersurfaces $\Sigma$ are affine spaces. That is, there exist corresponding real vector spaces $L_M$ and $L_\Sigma$ with transitive and free abelian group actions $L_M\times A_M\to A_M$ and $L_\Sigma\times A_\Sigma\to A_\Sigma$ respectively, written as addition ``$+$''. This allows to identify canonically all the tangent spaces $T_\phi A_M$ with $L_M$ and $T_\phi A_\Sigma$ with $L_\Sigma$ respectively. On hypersurfaces, the symplectic potential may then be seen as a map $\theta_\Sigma:A_\Sigma\times L_\Sigma\to\R$, linear in the second argument. We shall switch from here onwards to the notation $\theta_\Sigma(\varphi,\xi)$ instead of the previous notation $(\theta_\Sigma)_\varphi(\xi)$. Our second key assumption is that the symplectic potential is equivariant with respect to the affine structure in the following sense: There exists a bilinear form $[\cdot,\cdot]_\Sigma:L_\Sigma\times L_\Sigma\to\R$ such that
\begin{equation}
 \theta_\Sigma(\varphi+\xi,\phi)=\theta_\Sigma(\varphi,\phi)+[\xi,\phi]_\Sigma \qquad \forall \varphi\in A_\Sigma,\forall \xi,\phi\in L_\Sigma .
\end{equation}
This implies in turn that the symplectic structure is independent of the base point and may be viewed as an anti-symmetric bilinear map $\omega_\Sigma:L_\Sigma\times L_\Sigma\to\R$ given in terms of the symplectic potential as follows,
\begin{equation}
 \omega_\Sigma(\phi,\phi')=\frac{1}{2}[\phi,\phi']_\Sigma-\frac{1}{2}[\phi',\phi]_\Sigma \qquad\forall\phi,\phi'\in L_\Sigma .
\end{equation}
For a region $M$, the relation (\ref{eq:relspact}) can then be integrated to determine the action $S_M$ in terms of the symplectic potential $\theta_{\partial M}$, up to a constant,
\begin{equation}
 S_M(\eta)=S_M(\eta')-\frac{1}{2}\theta_{\partial M}(\eta,\eta-\eta')-\frac{1}{2}\theta_{\partial M}(\eta',\eta-\eta') \qquad\forall \eta,\eta'\in A_M .
\label{eq:propaact}
\end{equation}
For simplicity of notation we have omitted writing explicitly the composition with the map $a_M:A_M\to A_{\partial M}$ in the arguments of the symplectic potential.

\subsection{Ingredients from geometric quantization}
\label{sec:geomquant}

In order to construct the quantum state spaces associated to hypersurfaces we will use ingredients from geometric quantization. We thus proceed to give a lightning review of geometric quantization with special attention to the relevant case of \emph{holomorphic} or \emph{Kähler} quantization. We warn the reader that the following account is mostly based on thinking of phase space as a finite-dimensional manifold. Moreover, it is highly simplified and inaccurate in various respects. Nevertheless, it will suffice for our purposes. For a proper appreciation of geometric quantization we refer to standard text books such as \cite{Woo:geomquant}.

Geometric quantization of a classical phase space $A$ with symplectic two-form $\omega$ proceeds in two steps: A hermitian line bundle $B$, the \emph{prequantum bundle} is constructed over $A$, equipped with a connection $\nabla$ whose curvature is given by the symplectic form $\omega$. The prequantized Hilbert space $H$ is then given by square-integrable sections of $B$ with respect to a measure $\mu$ that is invariant under symplectic transformations. The inner product between sections $s',s$ is thus,
\begin{equation}
\langle s',s\rangle=\int (s'(\eta),s(\eta))_\eta\,\xd\mu(\eta),
\label{eq:pqip}
\end{equation}
where $(\cdot,\cdot)_\eta$ denotes the hermitian inner product on the fiber over $\eta\in A$. Note that a symplectic potential, i.e., a one-form $\theta$ over $A$ such that $\omega=\xd\theta$ gives rise to a trivialization of the bundle $B$ through the choice of a special section $s:A\to B$ that satisfies
\begin{equation}
 \nabla_X s=-\im\,\theta(X)\cdot s
\label{eq:trivsec}
\end{equation}
for all vector fields $X$ on $A$. Any other section of $B$ can then be obtained as $f s$, where $f$ is a complex valued function on $A$. We then have
\begin{equation}
 \nabla_X (f s)=(-\im\,\theta(X)\cdot f+\xd f(X))\, s .
\label{eq:trivid}
\end{equation}
 Moreover, by adjusting the overall normalization of $s$ if necessary we can arrange
\begin{equation}
 (s(\eta),s(\eta))_\eta=1 \qquad \forall \eta\in A .
\label{eq:normsec}
\end{equation}
The inner product (\ref{eq:pqip}) may then be written as,
\begin{equation}
\langle f's, f s\rangle=\int \overline{f'(\eta)} f(\eta)\,\xd\mu(\eta) .
\label{eq:trivip}
\end{equation}

While $H$ is usually ``too large'', the ``true'' Hilbert space of states $\cH$ is then obtained by a suitable restriction of $H$ through a \emph{polarization}. This is the second step. A polarization consists roughly of a choice of Lagrangian subspace $P_\eta$ of the complexified tangent space $T_\eta A^\C$ for each point $\eta\in A$. One then defines \emph{polarized sections} of $B$ to be those $s:B\to A$ satisfying
\begin{equation}
\nabla_{\overline{X}} s=0,
\end{equation}
where $X$ is a complex vector field valued at each point $\eta\in A$ in the polarized subspace $P_\eta\subseteq T_\eta A^\C$. The restriction of $H$ to the polarized sections yields the Hilbert space $\cH$.

In the holomorphic case, the polarization is induced by a complex structure $J_\eta$ on the tangent spaces $T_\eta A$, which is at the same time a symplectic transformation. Then, $\phi\mapsto \frac{1}{2}(\phi-\im J_\eta \phi)$ projects onto the polarized subspace $P_\eta\subseteq T_\eta A^\C$. At least locally, there exists then a \emph{Kähler potential} $K:A\to\R$ and an adapted complex symplectic potential $\Theta$ such that
\begin{equation}
 \Theta=-\im\sum_i \frac{\partial K}{\partial z_i}\xd z_i,
\end{equation}
where $\{z_i\}$ are local holomorphic coordinates with respect to the complex structure $J$. We can choose a (local) section $u$ of $B$ satisfying (\ref{eq:trivsec}) with respect to the complex one-form $\Theta$ to trivialize $B$. Then, general sections of $B$ can be obtained as $f u$ with $f$ a complex valued function on $A$. They satisfy the analogue of equation (\ref{eq:trivid}) with $\theta$ replaced by $\Theta$ and $s$ replaced by $u$. The point is that the subspace of polarized sections admits a simple description in terms of this trivialization. Namely, the polarized sections are now precisely the sections $f u$, where $f$ is a \emph{holomorphic} function on $A$. Since $\Theta$ is complex, the section $u$ cannot be normalized in analogy to (\ref{eq:normsec}). However, it can be related to the section $s$ that satisfies (\ref{eq:trivsec}) with respect to a given real symplectic potential $\theta$. Indeed, let $\alpha$ be the complex function on $A$ such that $u=\alpha s$. Then, we can use (\ref{eq:trivip}) to write the inner product on $\cH$ as follows,
\begin{equation}
 \langle f' u, f u\rangle=\int \overline{f'(\eta)} f(\eta)\, |\alpha(\eta)|^2\,\xd\mu(\eta) .
\label{eq:holip}
\end{equation}

As already mentioned the above account of geometric quantization is inaccurate in various respects. Nevertheless it is good enough to motivate our further discussion which will be limited to the case of affine field theory. Thus, we seek to quantize the space of solutions $A_\Sigma$ associated to a hypersurface $\Sigma$. The key additional ingredient apart from the classical data already described is a \emph{complex structure} on the tangent spaces of $A_\Sigma$. Since these tangent spaces are all canonically identified with $L_\Sigma$ and the symplectic structure is independent of the base point it will suffice to consider a single complex structure on $L_\Sigma$ as was the case in the treatment of linear field theory in \cite{Oe:holomorphic}. Thus, the complex structure is a linear map $J_\Sigma:L_\Sigma\to L_\Sigma$ satisfying $J_\Sigma^2=-\id_\Sigma$ and $\omega_\Sigma(J_\Sigma(\cdot), J_\Sigma(\cdot))=\omega_\Sigma(\cdot,\cdot)$. This gives rise to the symmetric bilinear form $g_\Sigma:L_\Sigma\times L_\Sigma\to\R$ by
\begin{equation}
 g_\Sigma(\phi,\eta)\defeq 2\omega_\Sigma(\phi,J_\Sigma \eta) \qquad\forall\phi,\eta\in L_\Sigma .
\label{eq:kmetric}
\end{equation}
We shall assume that this form is positive definite. The next step is to complete $L_\Sigma$ to a real Hilbert space with the inner product $g_\Sigma$. (We will continue to write $L_\Sigma$ for this completion.) It is then true that the sesquilinear form
\begin{equation}
\{\phi,\eta\}_\Sigma\defeq g_\Sigma(\phi,\eta)+2\im\omega_\Sigma(\phi,\eta) \qquad\forall\phi,\eta\in L_\Sigma
\label{eq:kcip}
\end{equation}
makes $L_\Sigma$ into a complex Hilbert space, where multiplication with $\im$ is given by applying $J_\Sigma$.

As discussed above, the complex structure $J_\Sigma$ defines a polarization and implies the existence of a Kähler potential. The Kähler potential is not unique, but a choice of base point $\eta\in A_\Sigma$ gives a natural definition of $K_\Sigma^\eta:A_\Sigma\to\R$ via
\begin{equation}
 K_\Sigma^\eta(\varphi)\defeq\frac{1}{2}g_\Sigma(\varphi-\eta,\varphi-\eta) .
\end{equation}
The adapted symplectic potential $\Theta_\Sigma^\eta:A_\Sigma\times L_\Sigma\to\C$ is then,
\begin{equation}
 \Theta_\Sigma^\eta(\varphi,\xi)=-\frac{\im}{2}\{\varphi-\eta,\xi\}_\Sigma .
\end{equation}
Recall that on the other hand we have the real symplectic potential $\theta_\Sigma:A_\Sigma\times L_\Sigma\to\R$. Suppose $s_\Sigma$ and $u_\Sigma^\eta$ are sections of the prequantum bundle $B_\Sigma$ over $A_\Sigma$ satisfying (\ref{eq:trivsec}) with respect to $\theta_\Sigma$ and $\Theta_\Sigma^\eta$ respectively. Moreover suppose that $s_\Sigma$ is normalized in the sense of (\ref{eq:normsec}). Let $\alpha_\Sigma^\eta:A_\Sigma\to\C$ be such that $u_\Sigma^\eta=\alpha_\Sigma^\eta s_\Sigma$. Then, it follows from (\ref{eq:trivid}) applied with $\theta_\Sigma$ on the one hand and with $\Theta_\Sigma^\eta$ on the other that,
\begin{equation}
 \xd\alpha_\Sigma^\eta=-\im\alpha_\Sigma^\eta(\Theta_\Sigma^\eta-\theta_\Sigma) .
\end{equation}
This determines $\alpha_\Sigma^\eta$ up to a constant factor, which is unimportant as it can be reabsorbed into the normalization of $u_\Sigma^\eta$. We set
\begin{equation}
\alpha_\Sigma^\eta(\varphi)\defeq\exp\left(\frac{\im}{2} \theta_\Sigma(\eta,\varphi-\eta) + \frac{\im}{2} \theta_\Sigma(\varphi,\varphi-\eta)-\frac{1}{4} g_\Sigma(\varphi-\eta,\varphi-\eta)\right) .
\label{eq:alphagq}
\end{equation}
We would then like to define the Hilbert space $\cH_\Sigma$ to consist of sections of $B$ that can be written as $f u_\Sigma^\eta$ with $f$ a holomorphic function on $A$. The inner product would be given by formula (\ref{eq:holip}), where $\mu$ is a probability measure on $A_\Sigma$ invariant under translations by elements of $L_\Sigma$. Indeed, if $A_\Sigma$ is finite-dimensional this immediately yields a nicely defined Hilbert space. However, in the more interesting case that $A_\Sigma$ is infinite-dimensional no such measure $\mu$ exists. Thinking of the factor $|\alpha|^2$ in (\ref{eq:holip}) as being part of a measure $\nu= |\alpha|^2\mu$ improves the situation. Still, no such measure $\nu$ exists on $A_\Sigma$. However, thinking of $\nu$ as living on $L_\Sigma$ rather than on $A_\Sigma$ and suitably extending to a larger space $\hat{L}_\Sigma$ does yield a well-defined measure. Such measures are well known, see e.g.\ \cite{GeVi:genf4} and an explicit construction suitable for the present setting was provided in \cite{Oe:holomorphic}. The latter will be used in Section~\ref{sec:quantization} to give a properly defined analogue of (\ref{eq:holip}).

\subsection{Ingredients from Feynman quantization}
\label{sec:feynquant}

Attempts to construct the amplitude maps associated to spacetime regions via quantization schemes that describe (time-)evolution through infinitesimal generators meet considerable difficulties. (Recall for example the difficulties in making the Tomonaga-Schwinger approach \cite{Tom:relwave,Sch:qed1} well defined.) In contrast, the Feynman path integral provides a conceptually much more satisfying approach to amplitudes. Of course, it comes with its own difficulties, but these do not show up in the simple setting of affine field theory considered here.

Recall in particular, that the combination of the Feynman path integral with the Schrödinger representation yields a rather direct construction of amplitude maps \cite{Oe:gbqft,Oe:kgtl}. To put this into the present context we recall that the Schrödinger representation may be seen as a particular case of geometric quantization with a real polarization. In the language of Section~\ref{sec:geomquant}, given a point $\eta$ in the space of solutions $A$, the polarized subspace $P_\eta$ of the complexified tangent space $T_\eta A^\C$ arises as the complexification of a real subspace $Q_\eta$ of the real tangent space $T_\eta A$. More specifically, in the case of the Schrödinger representation $Q_\eta$ is the subspace generated by the ``momenta'' $\partial\phi$ (in the notation of Section~\ref{sec:classft}). Fix $s$ to be the section of the prequantum bundle $B$ over $A$ satisfying (\ref{eq:trivsec}) with respect to the symplectic potential (\ref{eq:sympot}) as well as (\ref{eq:normsec}). Then, the polarized sections of $B$ are those that take the form $f s$, where $f$ is a complex function on $A$ that depends only on ``position'' coordinates $\phi$.

If $M$ is a spacetime region and $f s_{\partial M}$ a state in the Schrödinger polarized boundary Hilbert space, its amplitude is given heuristically by the Feynman path integral via
\begin{equation}
\rho_M(f s_{\partial M})=\int_{K_M} f(\zeta) \exp\left(\im S_M(\zeta)\right)\,\xd\mu(\zeta) ,
\label{eq:pathint}
\end{equation}
where $K_M$ is the space of field configurations in $M$ and $\mu$ is a measure on it that is invariant under symplectic transformations. Of course, no such measure exists and even the precise definition of the space $K_M$ may be unclear. As a first step to improve the situation we assume that there is a correspondence between field configuration data on the boundary and solutions in the interior, i.e., $K_M$ splits additively into $K_M=A_M\oplus K_M^0$, where $A_M$ is the space of solutions in $M$ while $K_M^0$ is the space of field configurations in $M$ that vanish on the boundary. Then, (\ref{eq:pathint}) may be rewritten as
\begin{equation}
\rho_M(f s_{\partial M})=\int_{A_M} f(\zeta) \left(\int_{K_M^0} \exp\left(\im S_M(\zeta+\Delta)\right)\,\xd\mu(\Delta)\right)\xd\mu(\zeta) .
\label{eq:pathint2}
\end{equation}

To further improve the situation we switch to the special case of affine field theory. The action $S_M$ is thus a polynomial of degree two on $K_M$ and by the variational principle we obtain $S_M(\zeta+\Delta)=S_M(\zeta)+F_M(\Delta)$ for $\zeta\in A_M$ and with $F_M$ some function. (In the case where $S_M$ is quadratic $F_M=S_M$.) This allows to factorize the inner integrand in (\ref{eq:pathint2}) and, discarding a normalization factor that only depends on $M$, to arrive at the expression
\begin{equation}
\rho_M(f s_{\partial M})=\int_{A_M} f(\zeta) \exp\left(\im S_M(\zeta)\right)\,\xd\mu(\zeta) .
\label{eq:pathint3}
\end{equation}
This is still ill-defined, but it turns out (Section~\ref{sec:ampl}) that the problem with the definition of the measure may be resolved in a manner similar to that indicated in the previous Section.

In the present paper, however, we take (\ref{eq:pathint3}) as a motivation for defining amplitudes by the same (rigorous equivalent of) formula (\ref{eq:pathint3}), but with the Schrödinger representation replaced by the holomorphic representation, discussed in the previous Section. That is, instead of $f$ being a function on field configurations it is taken to have the form $f=\tilde{f}\alpha_\Sigma^\eta$, where $\tilde{f}$ is a holomorphic function on $A_{\partial M}$. ($\eta\in A_M$ is a base point, the choice of which is irrelevant at this point.) As will be shown elsewhere, this replacement step can be justified rigorously. For purposes of the present paper we merely offer the partial justification that in both cases (Schrödinger and holomorphic) we are interpreting formula (\ref{eq:pathint3}) at least with respect to the very same trivialization of the prequantum bundle.

%% file: gaxioms.tex
\section{Axioms for classical and quantum field theory}

\subsection{Geometric data}
\label{sec:geomax}

In the previous section we have referred to regions and hypersurfaces in some fixed global spacetime. In contrast, from now on we will use a notion of spacetime in the spirit of topological quantum field theory, which is more abstract, but also more flexible. Nevertheless, a precise meaning is given to the concepts of \emph{region} and \emph{hypersurface}. While the setting we use is identical to that of \cite{Oe:holomorphic} we recall it briefly here for completeness.

Concretely, our geometric setting is the following: There is a fixed positive integer $d\in\N$, the \emph{dimension} of spacetime. We are given a collection of oriented topological manifolds of dimension $d$, possibly with boundary, that we call \emph{regions}. Furthermore, there is a collection of oriented topological manifolds without boundary of dimension $d-1$ that we call \emph{hypersurfaces}. All manifolds may only have finitely many connected components. When we want to emphasize explicitly that a given manifold is in one of those collections we also use the attribute \emph{admissible}. These collections satisfy the following requirements:
\begin{itemize}
\item Any connected component of a region or hypersurface is admissible.
\item Any finite disjoint union of regions or of hypersurfaces is admissible.
\item Any boundary of a region is an admissible hypersurface.
\item If $\Sigma$ is a hypersurface, then $\overline{\Sigma}$, denoting the same manifold with opposite orientation, is admissible.
\end{itemize}
It will turn out to be convenient to also introduce \emph{empty regions}. An empty region is topologically simply a hypersurface, but thought of as an infinitesimally thin region. Concretely, the empty region associated with a hypersurface $\Sigma$ will be denoted by $\hat{\Sigma}$ and its boundary is defined to be the disjoint union $\partial \hat{\Sigma}=\Sigma\cup\overline{\Sigma}$. There is one empty region for each hypersurface (forgetting its orientation). When an explicit distinction is desirable we refer to the previously defined regions as \emph{regular regions}.

There is also a notion of \emph{gluing} of regions. Suppose we are given a region $M$ with its boundary a disjoint union $\partial M=\Sigma_1\cup\Sigma\cup\overline{\Sigma'}$, where $\Sigma'$ is a copy of $\Sigma$. ($\Sigma_1$ may be empty.) Then, we may obtain a new manifold $M_1$ by gluing $M$ to itself along $\Sigma,\overline{\Sigma'}$. That is, we identify the points of $\Sigma$ with corresponding points of $\Sigma'$ to obtain $M_1$. The resulting manifold $M_1$ might be inadmissible, in which case the gluing is not allowed.

Depending on the theory one wants to model, the manifolds may carry additional structure such as for example a differentiable structure or a metric. This has to be taken into account in the gluing and will modify the procedure as well as its possibility in the first place. Our description above is merely meant as a minimal one. Moreover, there might be important information present in different ways of identifying the boundary hypersurfaces that are glued. Such a case can be incorporated into our present setting by encoding this information explicitly through suitable additional structure on the manifolds.

For brevity we shall refer to a collection of regions and hypersurfaces with the properties given above as a \emph{spacetime system}. A spacetime system can be induced from a global spacetime manifold by taking suitable submanifolds. (This setting was termed a \emph{global background} in \cite{Oe:gbqft}.) On the other hand, a spacetime system may arise by considering regions as independent pieces of spacetime that are not a priori embedded into any global manifold. Indeed, depending on the context, it might be physically undesirable to assume knowledge of, or even existence of, a fixed global spacetime structure.

%% file: classdata.tex
\subsection{Classical data}
\label{sec:classdata}

Given a spacetime system, the considerations of Section~\ref{sec:ingredients} motivate the following axiomatic definition of a \emph{classical affine field theory}. At the same time these axioms provide a natural generalization of the respective axioms presented in \cite{Oe:holomorphic} for the case of linear field theory.

\begin{itemize}
\item[(C1)] Associated to each hypersurface $\Sigma$ is a complex separable Hilbert space $L_\Sigma$ and an affine space $A_\Sigma$ over $L_\Sigma$ with the induced topology. The latter means that there is a transitive and free abelian group action $L_\Sigma\times A_\Sigma\to A_\Sigma$ which we denote by $(\phi,\eta)\mapsto \phi+\eta$. The inner product in $L_\Sigma$ is denoted by $\{\cdot ,\cdot\}_\Sigma$. We also define $g_\Sigma(\cdot,\cdot)\defeq \Re\{\cdot ,\cdot\}_\Sigma$ and $\omega_\Sigma(\cdot,\cdot)\defeq \frac{1}{2}\Im\{\cdot ,\cdot\}_\Sigma$ and denote by $J_\Sigma:L_\Sigma\to L_\Sigma$ the scalar multiplication with $\im$ in $L_\Sigma$. Moreover we suppose there are continuous maps $\theta_\Sigma:A_\Sigma\times L_\Sigma\to\R$ and $[\cdot,\cdot]_\Sigma:L_\Sigma\times L_\Sigma\to\R$ such that $\theta_\Sigma$ is real linear in the second argument, $[\cdot,\cdot]_\Sigma$ is real bilinear, and both structures are compatible via
\begin{equation}
[\phi,\phi']_\Sigma+\theta_\Sigma(\eta,\phi')=\theta_\Sigma(\phi+\eta,\phi')\qquad\forall \eta\in A_\Sigma, \forall \phi,\phi'\in L_\Sigma .
\end{equation}
Finally we require
\begin{equation}
 \omega_\Sigma(\phi,\phi')=\frac{1}{2} [\phi,\phi']_\Sigma-\frac{1}{2} [\phi',\phi]_\Sigma
\qquad \forall  \phi,\phi'\in L_\Sigma .
\end{equation}
\item[(C2)] Associated to each hypersurface $\Sigma$ there is a homeomorphic involution $A_\Sigma\to A_{\overline{\Sigma}}$ and a compatible conjugate linear involution $L_\Sigma\to L_{\overline\Sigma}$ under which the inner product is complex conjugated. We will not write these maps explicitly, but rather think of $A_\Sigma$ as identified with $A_{\overline{\Sigma}}$ and $L_\Sigma$ as identified with $L_{\overline\Sigma}$. Then, $\{\phi',\phi\}_{\overline{\Sigma}}=\overline{\{\phi',\phi\}_\Sigma}$ and we also require $\theta_{\overline{\Sigma}}(\eta,\phi)=-\theta_\Sigma(\eta,\phi)$ and $[\phi,\phi']_{\overline{\Sigma}}=-[\phi,\phi']_\Sigma$ for all $\phi,\phi'\in L_\Sigma$ and $\eta\in A_\Sigma$.
\item[(C3)] Suppose the hypersurface $\Sigma$ decomposes into a disjoint
  union of hypersurfaces $\Sigma=\Sigma_1\cup\cdots\cup\Sigma_n$. Then,
  there is a homeomorphism $A_{\Sigma_1}\times\dots\times A_{\Sigma_n}\to A_\Sigma$ and a compatible isometric isomorphism of complex Hilbert spaces
  $L_{\Sigma_1}\oplus\cdots\oplus L_{\Sigma_n}\to L_\Sigma$. Moreover, these maps satisfy obvious associativity conditions. We will not write these maps explicitly, but rather think of them as identifications. Also, $\theta_\Sigma=\theta_{\Sigma_1}+\dots+\theta_{\Sigma_n}$ and $[\cdot,\cdot]_\Sigma=[\cdot,\cdot]_{\Sigma_1}+\dots+[\cdot,\cdot]_{\Sigma_n}$.
\item[(C4)] Associated to each region $M$ is a real vector space $L_M$ and an affine space $A_M$ over $L_M$. Also, there is a map $S_M:A_M\to\R$.
\item[(C5)] Associated to each region $M$ there is a map $a_M:A_M\to A_{\partial M}$ and a compatible linear map of real vector spaces $r_M:L_M\to L_{\partial M}$. We denote by $A_{\tilde{M}}$ the image of $A_M$ under $a_M$ and by $L_{\tilde{M}}$ the image of $L_M$ under $r_M$. $L_{\tilde{M}}$ is a closed Lagrangian subspace of the real Hilbert space $L_{\partial M}$ with respect to the symplectic form $\omega_{\partial M}$. We often omit the explicit mention of the maps $a_M$ and $r_M$. We also require $S_M(\eta)=S_M(\eta')$ if $a_M(\eta)=a_M(\eta')$, and
\begin{equation}
S_M(\eta)=S_M(\eta')-\frac{1}{2}\theta_{\partial M}(\eta,\eta-\eta')-\frac{1}{2}\theta_{\partial M}(\eta',\eta-\eta')\qquad\forall\eta,\eta'\in A_M .
\label{eq:actsympot}
\end{equation}
\item[(C6)] Let $M_1$ and $M_2$ be regions and $M\defeq M_1\cup M_2$ be their disjoint union. Then, there is a bijection $A_{M_1}\times A_{M_2}\to A_M$ and a compatible isomorphism of real vector spaces $L_{M_1}\oplus L_{M_2}\to L_M$ such that $a_M=a_{M_1}\times a_{M_2}$ and $r_M=r_{M_1}\times r_{M_2}$. Moreover, these maps satisfy obvious associativity conditions. Hence, we can think of them as identifications and omit their explicit mention in the following. We also require $S_M=S_{M_1}+S_{M_2}$.
\item[(C7)] Let $M$ be a region with its boundary decomposing as a disjoint union $\partial M=\Sigma_1\cup\Sigma\cup \overline{\Sigma'}$, where $\Sigma'$ is a copy of $\Sigma$. Let $M_1$ denote the gluing of $M$ to itself along $\Sigma,\overline{\Sigma'}$ and suppose that $M_1$ is a region. Note $\partial M_1=\Sigma_1$. Then, there is an injective map $a_{M;\Sigma,\overline{\Sigma'}}:A_{M_1}\toi A_{M}$ and a compatible injective linear map $r_{M;\Sigma,\overline{\Sigma'}}:L_{M_1}\toi L_{M}$ such that
\begin{equation}
 A_{M_1}\toi A_{M}\rightrightarrows A_\Sigma \qquad L_{M_1}\toi L_{M}\rightrightarrows L_\Sigma
\label{eq:exsintbdy}
\end{equation}
are exact sequences. Here, for the first sequence, the arrows on the right hand side are compositions of the map $a_M$ with the projections of $A_{\partial M}$ to $A_\Sigma$ and $A_{\overline{\Sigma'}}$ respectively (the latter identified with $A_\Sigma$). For the second sequence the arrows on the right hand side are compositions of the map $r_M$ with the projections of $L_{\partial M}$ to $L_\Sigma$ and $L_{\overline{\Sigma'}}$ respectively (the latter identified with $L_\Sigma$). 
We also require $S_{M_1}=S_M\circ a_{M;\Sigma,\overline{\Sigma'}}$.
Moreover, the following diagrams commute, where the bottom arrows are the projections.
\begin{equation}
\xymatrix{
  A_{M_1} \ar[rr]^{a_{M;\Sigma,\overline{\Sigma'}}} \ar[d]_{a_{M_1}} & & A_{M} \ar[d]^{a_{M}}\\
  A_{\partial M_1}  & & A_{\partial M} \ar[ll]} \qquad
\xymatrix{
  L_{M_1} \ar[rr]^{r_{M;\Sigma,\overline{\Sigma'}}} \ar[d]_{r_{M_1}} & & L_{M} \ar[d]^{r_{M}}\\
  L_{\partial M_1}  & & L_{\partial M} \ar[ll]}
\end{equation}
\end{itemize}

In the spirit of Section~\ref{sec:affine}, the spaces $A_M$ and $A_\Sigma$ should be thought of as spaces of classical solutions in $M$ or near $\Sigma$. Correspondingly the spaces $L_M$ and $L_\Sigma$ should be thought of as their tangent spaces. Since $A_M$ and $A_\Sigma$ are affine we can naturally identify the tangent spaces at different points so that we do not need to distinguish them. Moreover, we assume the Hilbert space structure on the tangent spaces to be invariant under this identification. Thus, compared to the setting in \cite{Oe:holomorphic}, where the spaces of solutions where assumed vector spaces, each axiom contains now corresponding statements for both types of spaces, $A$ and $L$, as well as a statement of their compatibility. The latter is always supposed to mean that given the commuting diagrams expressing a certain property for $A$ and $L$ separately, combining these diagrams with the action diagrams $L\times A\to A$ yields a commuting diagram. Note also that forgetting the spaces $A$ as well as the structures $S$, $\theta$ and $[\cdot,\cdot]$, the axioms (C1)--(C7) strictly reduce to those given in \cite{Oe:holomorphic}. One may also remark that the axioms present quite some redundancy. For example certain properties of the spaces $A$ together with compatibility imply certain properties of the spaces $L$ and vice versa. However, the explicit form of the axioms was motivated more by conceptual simplicity and comparability with \cite{Oe:holomorphic} rather than by minimality.

We recall the following basic fact from \cite{Oe:holomorphic}:

\begin{lem}
\label{lem:decis}
Let $M$ be a region. Then, $L_{\partial M}$ understood as a real Hilbert space decomposes into an orthogonal direct sum $L_{\partial M}=L_{\tilde{M}}\oplus J_{\partial M} L_{\tilde{M}}$.
\end{lem}

We will use the notation $\phi=\phi^{\mathrm{R}}+J_{\partial M} \phi^{\mathrm{I}}$ for this decomposition, where $\phi\in L_{\partial M}$, $\phi^{\mathrm{R}},\phi^{\mathrm{I}}\in L_{\tilde{M}}$. There is a similar decomposition for elements of $A_{\partial M}$ given by the following Lemma. 

\begin{lem}
\label{lem:decbdyas}
Let $M$ be a region. Then, $A_{\partial M}$ decomposes into a generalized direct sum $A_{\partial M}=A_{\tilde{M}}\oplus J_{\partial M} L_{\tilde{M}}$.
\end{lem}
\begin{proof}
Let $\varphi\in A_{\partial M}$. We first show that there exists a decomposition $\varphi=\varphi^{\mathrm{R}}+J_{\partial M} \varphi^{\mathrm{I}}$ with $\varphi^{\mathrm{R}}\in A_{\tilde{M}}$ and $\varphi^{\mathrm{I}}\in L_{\tilde{M}}$ and then proceed to show its uniqueness. Fix $\eta\in A_{\tilde{M}}$. Then $\phi\defeq \varphi-\eta$ is element of $L_{\partial M}$ and thus decomposes as $\phi=\phi^{\mathrm{R}}+J_{\partial M} \phi^{\mathrm{I}}$ with $\phi^{\mathrm{R}},\phi^{\mathrm{I}}\in L_{\tilde{M}}$ according to Lemma~\ref{lem:decis}. It is then easy to see that setting $\varphi^{\mathrm{R}}=\phi^{\mathrm{R}}+\eta$ and $\varphi^{\mathrm{I}}=\phi^{\mathrm{I}}$ yields the desired decomposition. Suppose we are given two decompositions of the required form, $\varphi=\varphi_1^{\mathrm{R}}+J_{\partial M} \varphi_1^{\mathrm{I}}=\varphi_2^{\mathrm{R}}+J_{\partial M} \varphi_2^{\mathrm{I}}$. Their difference is
$0=\varphi_1^{\mathrm{R}}-\varphi_2^{\mathrm{R}}+J_{\partial M} (\varphi_1^{\mathrm{I}}-\varphi_2^{\mathrm{I}})$. But $\varphi_1^{\mathrm{R}}-\varphi_2^{\mathrm{R}}\in L_{\tilde{M}}$ and $\varphi_1^{\mathrm{I}}-\varphi_2^{\mathrm{I}}\in L_{\tilde{M}}$ so the latter amounts to a decomposition of $0\in L_{\partial M}$ in the sense of Lemma~\ref{lem:decis}. Uniqueness implies then $0=\varphi_1^{\mathrm{R}}-\varphi_2^{\mathrm{R}}$ and $0=\varphi_1^{\mathrm{I}}-\varphi_2^{\mathrm{I}}$.
\end{proof}

%% file: caxioms.tex
\subsection{Core axioms of the GBF}
\label{sec:coreaxioms}

A quantum (field) theory is encoded in the GBF by assigning ``algebraic'' data to the geometric data of a spacetime system, again in the spirit of topological quantum field theory. More concretely, Hilbert spaces are assigned to hypersurfaces and amplitude maps to regions. This is made precise in the following list of core axioms. This list is essentially identical to that given in \cite{Oe:holomorphic} and included here for completeness. We refer to the cited paper for further explanations. For brevity we call a theory satisfying these axioms for a given spacetime system a \emph{general boundary quantum field theory} on the spacetime system.

\begin{itemize}
\item[(T1)] Associated to each hypersurface $\Sigma$ is a complex
  separable Hilbert space $\cH_\Sigma$, called the \emph{state space} of
  $\Sigma$. We denote its inner product by
  $\langle\cdot,\cdot\rangle_\Sigma$.
\item[(T1b)] Associated to each hypersurface $\Sigma$ is a conjugate linear
  isometry $\iota_\Sigma:\cH_\Sigma\to\cH_{\overline{\Sigma}}$. This map
  is an involution in the sense that $\iota_{\overline{\Sigma}}\circ\iota_\Sigma$
  is the identity on  $\cH_\Sigma$.
\item[(T2)] Suppose the hypersurface $\Sigma$ decomposes into a disjoint
  union of hypersurfaces $\Sigma=\Sigma_1\cup\cdots\cup\Sigma_n$. Then,
  there is an isometric isomorphism of Hilbert spaces
  $\tau_{\Sigma_1,\dots,\Sigma_n;\Sigma}:\cH_{\Sigma_1}\ctens\cdots\ctens\cH_{\Sigma_n}\to\cH_\Sigma$.
  The composition of the maps $\tau$ associated with two consecutive
  decompositions is identical to the map $\tau$ associated to the
  resulting decomposition.
\item[(T2b)] The involution $\iota$ is compatible with the above
  decomposition. That is,
  $\tau_{\overline{\Sigma}_1,\dots,\overline{\Sigma}_n;\overline{\Sigma}}
  \circ(\iota_{\Sigma_1}\ctens\cdots\ctens\iota_{\Sigma_n}) 
  =\iota_\Sigma\circ\tau_{\Sigma_1,\dots,\Sigma_n;\Sigma}$.
\item[(T4)] Associated with each region $M$ is a linear map
  from a dense subspace $\cH_{\partial M}^\ds$ of the state space
  $\cH_{\partial M}$ of its boundary $\partial M$ (which carries the
  induced orientation) to the complex
  numbers, $\rho_M:\cH_{\partial M}^\ds\to\C$. This is called the
  \emph{amplitude} map.
\item[(T3x)] Let $\Sigma$ be a hypersurface. The boundary $\partial\hat{\Sigma}$ of the associated empty region $\hat{\Sigma}$ decomposes into the disjoint union $\partial\hat{\Sigma}=\overline{\Sigma}\cup\Sigma'$, where $\Sigma'$ denotes a second copy of $\Sigma$. Then, $\tau_{\overline{\Sigma},\Sigma';\partial\hat{\Sigma}}(\cH_{\overline{\Sigma}}\tens\cH_{\Sigma'})\subseteq\cH_{\partial\hat{\Sigma}}^\ds$. Moreover, $\rho_{\hat{\Sigma}}\circ\tau_{\overline{\Sigma},\Sigma';\partial\hat{\Sigma}}$ restricts to a bilinear pairing $(\cdot,\cdot)_\Sigma:\cH_{\overline{\Sigma}}\times\cH_{\Sigma'}\to\C$ such that $\langle\cdot,\cdot\rangle_\Sigma=(\iota_\Sigma(\cdot),\cdot)_\Sigma$.
\item[(T5a)] Let $M_1$ and $M_2$ be regions and $M\defeq M_1\cup M_2$ be their disjoint union. Then $\partial M=\partial M_1\cup \partial M_2$ is also a disjoint union and $\tau_{\partial M_1,\partial M_2;\partial M}(\cH_{\partial M_1}^\ds\tens \cH_{\partial M_2}^\ds)\subseteq \cH_{\partial M}^\ds$. Then, for all $\psi_1\in\cH_{\partial M_1}^\ds$ and $\psi_2\in\cH_{\partial M_2}^\ds$,
\begin{equation}
 \rho_{M}\circ\tau_{\partial M_1,\partial M_2;\partial M}(\psi_1\tens\psi_2)= \rho_{M_1}(\psi_1)\rho_{M_2}(\psi_2) .
\end{equation}
\item[(T5b)] Let $M$ be a region with its boundary decomposing as a disjoint union $\partial M=\Sigma_1\cup\Sigma\cup \overline{\Sigma'}$, where $\Sigma'$ is a copy of $\Sigma$. Let $M_1$ denote the gluing of $M$ with itself along $\Sigma,\overline{\Sigma'}$ and suppose that $M_1$ is a region. Note $\partial M_1=\Sigma_1$. Then, $\tau_{\Sigma_1,\Sigma,\overline{\Sigma'};\partial M}(\psi\tens\xi\tens\iota_\Sigma(\xi))\in\cH_{\partial M}^\ds$ for all $\psi\in\cH_{\partial M_1}^\ds$ and $\xi\in\cH_\Sigma$. Moreover, for any ON-basis $\{\xi_i\}_{i\in I}$ of $\cH_\Sigma$, we have for all $\psi\in\cH_{\partial M_1}^\ds$,
\begin{equation}
 \rho_{M_1}(\psi)\cdot c(M;\Sigma,\overline{\Sigma'})
 =\sum_{i\in I}\rho_M\circ\tau_{\Sigma_1,\Sigma,\overline{\Sigma'};\partial M}(\psi\tens\xi_i\tens\iota_\Sigma(\xi_i)),
\label{eq:glueax1}
\end{equation}
where $c(M;\Sigma,\overline{\Sigma'})\in\C\setminus\{0\}$ is called the \emph{gluing anomaly factor} and depends only on the geometric data.
\end{itemize}

%% file: quantization.tex
\section{Quantization}
\label{sec:quantization}

In this section we describe a quantization prescription that produces for a given classical affine field theory (satisfying the axioms of Section~\ref{sec:classdata}) on a spacetime system a general boundary quantum field theory on the same spacetime system. In particular, we rigorously prove that the produced theory satisfies the core axioms of the GBF as presented in Section~\ref{sec:coreaxioms}.

\subsection{State Spaces}
\label{sec:sspaces}

As explained in \cite{Oe:holomorphic}, the inner product $\frac{1}{2}\{\cdot,\cdot\}_\Sigma$ on the complex Hilbert space $L_\Sigma$ defines a Gaussian measure $\nu_\Sigma$ on the space $\hat{L}_\Sigma$. Here, $\hat{L}_\Sigma$ is the algebraic dual of the topological dual of $L_\Sigma$ so that there is a natural inclusion $L_\Sigma\toi \hat{L}_\Sigma$. Recall furthermore that the square-integrable holomorphic functions on $\hat{L}_\Sigma$ form a separable complex Hilbert space $\rH^2(\hat{L}_\Sigma,\nu_\Sigma)$, whose elements are uniquely determined by their values on the subspace $L_\Sigma$ (Theorem~3.18 of \cite{Oe:holomorphic}). We denote the complex vector space of functions on $L_\Sigma$ that arise as restrictions of elements in $\rH^2(\hat{L}_\Sigma,\nu_\Sigma)$ by $\rH^2_\Sigma$. Obviously, $\rH^2_\Sigma$ inherits the inner product of $\rH^2(\hat{L}_\Sigma,\nu_\Sigma)$, making it naturally isomorphic to that space as a complex Hilbert space. Note that the elements of $\rH^2_\Sigma$ are in particular continuous functions on $L_\Sigma$.

Denote the algebra of complex valued continuous functions on $A_\Sigma$ by $\cont_\Sigma$. We define the Hilbert space $\cH_\Sigma$ associated to the hypersurface $\Sigma$ as a certain subspace of $\cont_\Sigma$ as follows. Fix a \emph{base point} $\eta\in A_\Sigma$ and define the following element of $\cont_\Sigma$, motivated by (\ref{eq:alphagq}),
\begin{equation}
\alpha_{\Sigma}^\eta(\varphi)\defeq\exp\left(\frac{\im}{2} \theta_\Sigma(\eta,\varphi-\eta) + \frac{\im}{2} \theta_\Sigma(\varphi,\varphi-\eta)-\frac{1}{4} g_\Sigma(\varphi-\eta,\varphi-\eta)\right) .
\label{eq:defalpha}
\end{equation}
Now, define $\cH_\Sigma$ as the subspace of $\cont_\Sigma$ of elements $\psi$ that take the form
\begin{equation}
 \psi(\varphi)=\chi^\eta(\varphi-\eta)\alpha_{\Sigma}^\eta(\varphi) ,
\label{eq:wfbase}
\end{equation}
where $\chi^\eta\in\rH^2_\Sigma$. Moreover, we define the inner product on $\cH_\Sigma$ as follows,\footnote{Here and in the following elements of $\rH^2_\Sigma$ appearing in an integral should be thought of as representing the respective elements of $\rH^2(\hat{L},\nu_\Sigma)$.}
\begin{equation}
\langle\psi',\psi\rangle_\Sigma = \int_{\hat{L}_\Sigma} \chi^\eta\overline{{\chi'}^\eta}\,\xd\nu_\Sigma .
\label{eq:hsip}
\end{equation}
Clearly, $\cH_\Sigma$ becomes a complex separable Hilbert space in this way, which is moreover naturally isomorphic to $\rH^2_\Sigma$. Moreover, it turns out that the definition is independent of the choice of base point.

\begin{lem}
\label{lem:hindepbp}
The above definition of $\cH_\Sigma$ is independent of the choice of base point.
\end{lem}
\begin{proof}
Fix $\eta,\tilde{\eta}\in A_\Sigma$. Straightforward computation yields,
\begin{multline}
\frac{\alpha_{\Sigma}^\eta(\varphi)}{\alpha_{\Sigma}^{\tilde{\eta}}(\varphi)}
= \exp\left(\frac{1}{2}\{\eta-\tilde{\eta},\varphi-\tilde{\eta}\}_\Sigma
 -\frac{1}{4}g_\Sigma(\eta-\tilde{\eta},\eta-\tilde{\eta})\right.\\
\left. -\frac{\im}{2}\theta_\Sigma(\eta,\eta-\tilde{\eta})
 -\frac{\im}{2}\theta_\Sigma(\tilde{\eta},\eta-\tilde{\eta})\right)
\label{eq:alphaquot}
\end{multline}
Thus, suppose we have $\psi\in\cont_\Sigma$ decomposed as in (\ref{eq:wfbase}) with respect to the base point $\eta$. We equate this to a decomposition with respect to the base point $\tilde{\eta}$,
\begin{equation}
 \chi^\eta(\varphi-\eta)\alpha_{\Sigma}^\eta(\varphi)
 =\chi^{\tilde{\eta}}(\varphi-\tilde{\eta})\alpha_{\Sigma}^{\tilde{\eta}}(\varphi) .
\end{equation}
Using (\ref{eq:alphaquot}) we obtain,
\begin{multline}
 \chi^{\tilde{\eta}}(\phi)=\chi^\eta(\phi+\tilde{\eta}-\eta)
 \exp\left(\frac{1}{2}\{\eta-\tilde{\eta},\phi\}_\Sigma
 -\frac{1}{4}g_\Sigma(\eta-\tilde{\eta},\eta-\tilde{\eta})\right.\\
\left. -\frac{\im}{2}\theta_\Sigma(\eta,\eta-\tilde{\eta})
 -\frac{\im}{2}\theta_\Sigma(\tilde{\eta},\eta-\tilde{\eta})\right)
\label{eq:chitrans}
\end{multline}
Note that the inner product $\{\cdot,\cdot\}_\Sigma$ is holomorphic in its second argument, so the exponential expression in (\ref{eq:chitrans}) is holomorphic in $\phi$. On the other hand $\chi^\eta$ is holomorphic by assumption and so is thus the composition of $\chi^\eta$ with a translation. Thus $\chi^{\tilde{\eta}}$ is holomorphic, being the product of holomorphic functions. Proposition~3.11 of \cite{Oe:holomorphic} with $(\cdot,\cdot)=\frac{1}{2} g_\Sigma(\cdot,\cdot)$, $p=2$, $f=\chi^{\eta}$ and $x=\tilde{\eta}-\eta$, yields that the extension of
\begin{equation}
\phi\mapsto \chi^{\eta}(\phi+\tilde{\eta}-\eta)\exp\left(-\frac{1}{4}g_\Sigma(2\phi+\tilde{\eta}-\eta,\tilde{\eta}-\eta)\right)
\label{eq:shiftchi}
\end{equation}
is square-integrable on $(\hat{L}_\Sigma,\nu_\Sigma)$. So this function is in $\rH^2(\hat{L}_\Sigma,\nu_\Sigma)$. On the other hand (\ref{eq:chitrans}) and (\ref{eq:shiftchi}) differ only by a constant factor, so the extension of $\chi^{\tilde{\eta}}$ is also in $\rH^2(\hat{L}_\Sigma,\nu_\Sigma)$. That is, $\chi^{\tilde{\eta}}\in\rH^2_\Sigma$. This already shows that $\cH_\Sigma$ as a subspace of $\cont_\Sigma$ is independent of the choice of base point.

It remains to show that the inner product (\ref{eq:hsip}) is also invariant under choice of base point. For two elements $\psi,\psi'\in\cH_\Sigma$ decompose as above with respect to two different base points $\eta,\tilde{\eta}\in A_\Sigma$. Then,
\begin{align*}
&\int_{\hat{L}_\Sigma} \chi^{\tilde{\eta}}(\phi)\overline{{\chi'}^{\tilde{\eta}}(\phi)}\,\xd\nu_\Sigma(\phi)\\ 
&=\int_{\hat{L}_\Sigma} \chi^{\eta}(\phi+\tilde{\eta}-\eta)\overline{{\chi'}^{\eta}(\phi+\tilde{\eta}-\eta)}\exp\left(-\frac{1}{2}g_\Sigma(2\phi+\tilde{\eta}-\eta,\tilde{\eta}-\eta)\right)\xd\nu_\Sigma(\phi) \\
&=\int_{\hat{L}_\Sigma} \chi^{\eta}(\phi)\overline{{\chi'}^{\eta}(\phi)}\,\xd\nu_\Sigma(\phi) .
\end{align*}
The second equality here is another consequence of Proposition~3.11 of \cite{Oe:holomorphic}, applied as above. This completes the proof.
\end{proof}

Heuristically, the definition of the inner product (\ref{eq:hsip}) is motivated by the manifestly base point independent expression
\begin{equation}
\langle\psi',\psi\rangle_\Sigma = `` \int_{A_\Sigma}  \overline{\psi'(\eta)} \psi(\eta)\,\xd\mu_\Sigma(\eta) ``,
\end{equation}
where $\mu_\Sigma$ stands for a (non-existent) translation invariant measure on $A_\Sigma$, recall expression (\ref{eq:trivip}) in Section~\ref{sec:geomquant}.

We shall refer to the elements of $\cH_\Sigma$ also as \emph{wave functions}. Note that a function that is holomorphic on $L_\Sigma$ is anti-holomorphic on $L_{\overline{\Sigma}}$ and vice versa. Also, $\alpha_{\overline{\Sigma}}^{\eta}=\overline{\alpha_{\Sigma}^{\eta}}$. Thus, complex conjugation of wave functions yields a conjugate linear isomorphism $\iota_\Sigma:\cH_\Sigma\to\cH_{\overline{\Sigma}}$. For disjoint unions of hypersurfaces $\Sigma_1,\Sigma_2$ and $\eta_1\in A_{\Sigma_1}$, $\eta_2\in A_{\Sigma_2}$ we have $\alpha_{\Sigma_1\cup\Sigma_2}^{(\eta_1,\eta_2)}=\alpha_{\Sigma_1}^{\eta_1}\alpha_{\Sigma_2}^{\eta_2}$ and therefore naturally get $\cH_{\Sigma_1\cup\Sigma_2}=\cH_{\Sigma_1}\ctens\cH_{\Sigma_2}$, where the tensor product is the (completed) tensor product of Hilbert spaces. Thus, we have satisfied core axioms (T1), (T1b), (T2), (T2b) of Section~\ref{sec:coreaxioms}.

Recall from \cite{Oe:holomorphic} that the Hilbert space of states associated with a hypersurface $\Sigma$ for the linear space of solutions $L_\Sigma$ is precisely the space that we called $\rH^2_\Sigma$ above. Unsurprisingly, the choice of a base point in the affine space of solutions $A_\Sigma$ not only yields a natural identification of $L_\Sigma$ with $A_\Sigma$, but also yields a natural isomorphism between $\rH^2_\Sigma$ and $\cH_\Sigma$ via (\ref{eq:wfbase}) as described above. This will allow us to import many of the results of \cite{Oe:holomorphic} into the present setting.

\subsection{Coherent States}
\label{sec:cohstates}

As in the linear case, coherent states provide also in affine field theory a convenient and powerful tool in laying out the structure of the quantum theory. Indeed, choosing a base point, we can directly import the coherent states as presented in \cite{Oe:holomorphic}. Recall from \cite{Oe:holomorphic} that coherent states in the linear theory are indexed by elements of $L_\Sigma$. In particular, the coherent state $K_\xi$ for $\xi\in L_\Sigma$ is the element of $\rH^2_\Sigma$ given by,
\begin{equation}
 K_\xi(\phi)=\exp\left(\frac{1}{2}\{\xi, \phi\}_\Sigma\right)\qquad\forall \phi\in L_\Sigma .
\end{equation}
Choosing a base point $\eta\in A_\Sigma$ we will denote the element of $\cH_\Sigma$ corresponding to $K_\xi$ via (\ref{eq:wfbase}) as $K_{\xi}^\eta$,
\begin{equation}
K_{\xi}^\eta(\varphi)\defeq K_\xi(\varphi-\eta)\alpha_{\Sigma}^{\eta}(\varphi)\qquad\forall \varphi\in A_\Sigma .
\label{eq:deccohwf}
\end{equation}
It is preferable, however, to have an intrinsic concept of coherent state adapted to the affine setting. This should yield a coherent state associated to any element of $A_\Sigma$, and be independent of a choice of base point. To this end we recall that a key property of the coherent states is the \emph{reproducing property}. That is, taking the inner product of a coherent state with an arbitrary state yields the wave function of the state evaluated at the point corresponding to the coherent state. In the present context this leads to the following definition of an \emph{affine coherent state} $\hat{K}_\zeta\in\cH_\Sigma$ associated to the element $\zeta\in A_\Sigma$,
\begin{equation}
\hat{K}_\zeta(\varphi)\defeq\exp\left(\frac{\im}{2}\theta_\Sigma(\zeta,\varphi-\zeta)+\frac{\im}{2}\theta_\Sigma(\varphi,\varphi-\zeta)-\frac{1}{4}g_\Sigma(\varphi-\zeta,\varphi-\zeta)\right) .
\label{eq:defacoh}
\end{equation}
For $\eta\in A_\Sigma$ and $\xi\in L_\Sigma$ the relation between $\hat{K}_{\eta+\xi}$ and $K^\eta_\xi$ is merely a constant factor,
\begin{equation}
 \hat{K}_{\eta+\xi}=K^\eta_\xi \exp\left(-\im\theta_\Sigma(\eta,\xi)-\frac{\im}{2}[\xi,\xi]_\Sigma-\frac{1}{4}g_\Sigma(\xi,\xi)\right) .
\label{eq:relcohst}
\end{equation}
Note that the real part in the exponential precisely normalizes the state, so that $\hat{K}_{\eta+\xi}$ differs from the normalized version of $K^\eta_\xi$ merely by a constant phase.

Some basic properties of affine coherent states are easily deduced from properties of their linear counterparts, see Propositions~3.14 and 3.19 as well as Section~4.2 of \cite{Oe:holomorphic},
\begin{gather}
\langle \hat{K}_\zeta,\psi\rangle_\Sigma=\psi(\zeta),\\
\langle \hat{K}_{\zeta'},\hat{K}_{\zeta}\rangle_\Sigma=\exp\left(\frac{\im}{2}\theta_\Sigma(\zeta,\zeta'-\zeta)+\frac{\im}{2}\theta_\Sigma(\zeta',\zeta'-\zeta)-\frac{1}{4}g_\Sigma(\zeta'-\zeta,\zeta'-\zeta)\right) \label{eq:ipacoh},\\
\|\hat{K}_\zeta\|_2= 1,\\
 \langle\psi',\psi\rangle_\Sigma=\int_{\hat{L}_\Sigma} \langle\psi',\hat{K}_{\eta+\xi}\rangle_\Sigma
  \langle \hat{K}_{\eta+\xi},\psi\rangle_\Sigma\exp\left(\frac{1}{2}g_\Sigma(\xi,\xi)\right)\,
 \xd\nu_\Sigma(\xi).
\label{eq:cohcompl}
\end{gather}
Note that in the completeness relation (\ref{eq:cohcompl}) we use a base point $\eta\in A_\Sigma$ as we have a measure on $\hat{L}_\Sigma$, rather than on $\hat{A}_\Sigma$. However, the choice of base point is arbitrary as the left hand side does not depend on it.

When it is useful, we also indicate explicitly on which hypersurface a coherent state lives, e.g., we write $\hat{K}_{\Sigma,\zeta}$ to indicated this. Coherent states are compatible with the involutions $\iota_{\Sigma}:\cH_{\Sigma}\to\cH_{\overline{\Sigma}}$ in the obvious way,
\begin{equation}
\hat{K}_{\overline{\Sigma},\zeta}=\iota_{\Sigma}(\hat{K}_{\Sigma,\zeta}).
\label{eq:involcoh}
\end{equation} 
The coherent states are also compatible with decompositions of hypersurfaces in a simple way. Namely, for $(\zeta,\zeta')\in A_{\Sigma}\times A_{\Sigma'}$ we have
\begin{equation}
 \hat{K}_{\Sigma\cup\Sigma',(\zeta,\zeta')}=\hat{K}_{\Sigma,\zeta}\tens \hat{K}_{\Sigma',\zeta'} .
\label{eq:deccoh}
\end{equation}

\subsection{Amplitudes}
\label{sec:ampl}

Let $M$ be a region. Recall from \cite{Oe:holomorphic} that $g_{\partial M}$ viewed as the inner product on the real Hilbert space $L_{\tilde M}$ induces a measure $\nu_{\tilde{M}}$ on $\hat{L}_{\tilde{M}}\subseteq \hat{L}_{\partial M}$. This was used there to define the amplitude map on a dense subspace of the boundary Hilbert space. We shall proceed in a similar manner here. At first, we fix a base point $\eta\in A_{M}$. For a state $\psi\in\cH_\Sigma$ we consider the decomposition (\ref{eq:wfbase}). As explained previously, due to Theorem~3.18 of \cite{Oe:holomorphic} the element $\chi^\eta\in\rH^2_\Sigma$ uniquely extends to an element in $\rH^2(\hat{L}_{\partial M},\nu_{\partial M})$, which we shall also denote by $\chi^\eta$. Now, $\chi^\eta$ viewed as a map on $\hat{L}_{\tilde{M}}$ (we omit writing explicitly the composition with the map $r_M$ or its extension) may or may not be in $\cL^1(\hat{L}_{\tilde{M}},\nu_{\tilde{M}})$. If $\chi^\eta$ is integrable in this sense, we define the amplitude of the state $\psi$ as follows,
\begin{equation}
 \rho_M(\psi)\defeq\exp\left(\im S_M(\eta)\right)\int_{\hat{L}_{\tilde{M}}} \chi^\eta(\phi)\,\xd\nu_{\tilde{M}}(\phi) .
\label{eq:defampl}
\end{equation}
Our first task will be to show that this definition is independent of the choice of base point.

\begin{lem}
\label{lem:amplibp}
The above definition of $\rho_M(\psi)$ is independent of the base point.
\end{lem}
\begin{proof}
Fix $\eta,\tilde{\eta}\in A_M$. We find,
\begin{align}
&\exp\left(\im S_M(\eta)\right)\int_{\hat{L}_{\tilde{M}}} \chi^\eta(\phi)\,\xd\nu_{\tilde{M}}(\phi)\label{eq:aib1}\\
& = \exp\left(\im S_M(\eta)\right)\int_{\hat{L}_{\tilde{M}}} \chi^\eta(\phi+\tilde{\eta}-\eta)\exp\left(-\frac{1}{4}g_{\partial M}(2\phi+\tilde{\eta}-\eta,\tilde{\eta}-\eta)\right)\xd\nu_{\tilde{M}}(\phi)
\label{eq:aib2}\\
& = \exp\left(\im S_M(\eta)\right)\int_{\hat{L}_{\tilde{M}}} \chi^{\tilde{\eta}}(\phi)\nonumber\\
& \qquad \exp\left(\im\omega_{\partial M}(\phi,\tilde{\eta}-\eta)-\frac{\im}{2}\theta_{\partial M}(\eta,\tilde{\eta}-\eta)-\frac{\im}{2}\theta_{\partial M}(\tilde{\eta},\tilde{\eta}-\eta)\right)\xd\nu_{\tilde{M}}(\phi)\label{eq:aib3}\\
& = \exp\left(\im S_M(\tilde{\eta})\right)\int_{\hat{L}_{\tilde{M}}} \chi^{\tilde{\eta}}(\phi)\,\xd\nu_{\tilde{M}}(\phi)
\label{eq:aib4}
\end{align}
The equality between (\ref{eq:aib1}) and (\ref{eq:aib2}) follows from an application of Proposition~3.11 of \cite{Oe:holomorphic} with $(\cdot,\cdot)=\frac{1}{4} g_\Sigma(\cdot,\cdot)$, $p=1$, $f=\chi^{\eta}$ and $x=\tilde{\eta}-\eta$. The equality between (\ref{eq:aib2}) and (\ref{eq:aib3}) follows from (\ref{eq:chitrans}) in the proof of Lemma~\ref{lem:hindepbp}. Finally, the equality between (\ref{eq:aib3}) and (\ref{eq:aib4}) follows from equation (\ref{eq:actsympot}) of axiom (C5) together with the fact that $L_{\tilde{M}}$ is a Lagrangian subspace of $L_{\partial{M}}$ and hence $\omega_{\partial M}(\phi,\tilde{\eta}-\eta)=0$.
\end{proof}

As is easily seen, the expression (\ref{eq:defampl}) for the amplitude can be equivalently written as follows,
\begin{equation}
 \rho_M(\psi)=\int_{\hat{L}_{\tilde{M}}} \psi(\eta+\phi)\,\exp\left(\im S_M(\eta+\phi)+\frac{1}{4}g_{\partial M}(\phi,\phi)\right)\,\xd\nu_{\tilde{M}}(\phi) .
\end{equation}
Heuristically, this is suggested by the following manifestly base point independent formula
\begin{equation}
\rho_M(\psi) = `` \int_{A_{M}} \psi(\eta) \exp\left(\im S_M(\eta)\right)\,\xd\mu_M(\eta) ``,
\end{equation}
where $\mu_M$ stands for a (non-existent) translation invariant measure on $A_M$, recall expression (\ref{eq:pathint3}) in Section~\ref{sec:feynquant}.

As we shall see in a moment the subspace $\cH^\ds_{\partial M}$ of $\cH_{\partial M}$ of wave functions which are integrable contains at least all coherent states. Thus, $\cH^\ds_{\partial M}$ is dense in $\cH_{\partial M}$ by Proposition~3.15 of \cite{Oe:holomorphic}. Therefore, axiom (T4) is satisfied.

It is also clear that formula (\ref{eq:defampl}) satisfies axiom (T5a). Indeed, the measure for a disjoint union of regions $M_1$, $M_2$ is the product measure. Choosing base points $(\eta_1,\eta_2)\in A_{M_1}\times A_{M_2}=A_{M_1\cup M_2}$ one may then observe that the integral for $M_1\cup M_2$ coincides with the product of the respective integrals for $M_1$ and $M_2$. The corresponding factorization of the exponential pre-factor follows from the additivity of $S$ as exhibited in axiom (C6).

As in the linear case treated in \cite{Oe:holomorphic} it is possible to explicitly evaluate the amplitude on coherent states.
\begin{prop}
\label{prop:cohampl}
Let $\zeta\in A_{\partial M}$ and $\zeta=\zeta^{\mathrm{R}}+J_{\partial M} \zeta^{\mathrm{I}}$ be its decomposition with respect to the generalized direct sum $A_{\partial M}=A_{\tilde{M}}\oplus J_{\partial M} L_{\tilde{M}}$ according to Lemma~\ref{lem:decbdyas}. Then,
\begin{multline}
 \rho_M(\hat{K}_\zeta)=\\
\exp\left(\im S_M(\zeta^{\mathrm{R}})-\im\,\theta_{\partial M}(\zeta^{\mathrm{R}},J_{\partial M}\zeta^{\mathrm{I}})-\frac{\im}{2}[J_{\partial M}\zeta^{\mathrm{I}},J_{\partial M}\zeta^{\mathrm{I}}]_{\partial M}-\frac{1}{2} g_{\partial M}(\zeta^{\mathrm{I}},\zeta^{\mathrm{I}})\right) .
\label{eq:cohampl}
\end{multline}
\end{prop}
\begin{proof}
Fixing a base point $\eta\in A_{\tilde{M}}$ we decompose the coherent state wave function as in (\ref{eq:wfbase}),
\begin{equation}
\hat{K}_\zeta(\varphi)=\chi^\eta_\zeta(\varphi-\eta)\alpha_{\partial M}^\eta(\varphi) .
\end{equation}
Combining (\ref{eq:relcohst}) and (\ref{eq:deccohwf}) then yields
\begin{equation}
\chi^\eta_\zeta=K_{\zeta-\eta}\exp\left(-\im\,\theta_{\partial M}(\eta,\zeta-\eta)
 -\frac{\im}{2}[\zeta-\eta,\zeta-\eta]_{\partial{M}}-\frac{1}{4}g_{\partial{M}}(\zeta-\eta,\zeta-\eta)\right) .
\label{eq:chicoh}
\end{equation}
Proposition~4.2 of \cite{Oe:holomorphic} gives the value of the relevant integral,
\begin{multline}
\int_{\hat{L}_{\tilde{M}}} K_{\zeta-\eta}(\phi)\,\xd\nu_{\tilde{M}}(\phi)
=\exp\left(\frac{1}{4}g_{\partial M}\left((\zeta-\eta)^{\mathrm{R}},(\zeta-\eta)^{\mathrm{R}}\right)\right.\\
\left.-\frac{1}{4}g_{\partial M}\left((\zeta-\eta)^{\mathrm{I}},(\zeta-\eta)^{\mathrm{I}}\right)-\frac{\im}{2}g_{\partial M}\left((\zeta-\eta)^{\mathrm{R}},(\zeta-\eta)^{\mathrm{I}}\right)\right),
\end{multline}
where the decomposition of $L_{\partial M}$ according to Lemma~\ref{lem:decis} is used. The amplitude (\ref{eq:defampl}) is thus,
\begin{multline}
 \rho_M(\hat{K}_\zeta)=\exp\left(\im S_M(\eta)-\im\,\theta_{\partial M}(\eta,\zeta-\eta) -\frac{\im}{2}[\zeta-\eta,\zeta-\eta]_{\partial{M}}\right.\\
\left.-\frac{1}{2}g_{\partial M}\left((\zeta-\eta)^{\mathrm{I}},(\zeta-\eta)^{\mathrm{I}}\right)-\frac{\im}{2}g_{\partial M}\left((\zeta-\eta)^{\mathrm{R}},(\zeta-\eta)^{\mathrm{I}}\right)\right) .
\end{multline}
Straightforward computation using the decomposition of $A_{\partial M}$ according to Lemma~\ref{lem:decbdyas}, the formula (\ref{eq:actsympot}) and other basic identities leads to (\ref{eq:cohampl}).
\end{proof}

Recall that a simple, but compelling physical interpretation of the linear analogue of the amplitude formula (\ref{eq:cohampl}) was put forward in \cite{Oe:holomorphic}. Essentially this same interpretation extends to the present affine setting as follows. If we think in classical terms, the component $\zeta^{\mathrm{R}}$ of the boundary solution $\zeta$ can be continued consistently to the interior and is hence classically allowed. The component $J_{\partial M}\zeta^{\mathrm{I}}$ does not possess such a continuation and is hence classically forbidden. This is reflected precisely in equation (\ref{eq:cohampl}). If the classically forbidden component is not present, the amplitude has unit modulus. Its phase is irrelevant for probabilities or expectation values related to measurements in $M$. On the other hand, the presence of a classically forbidden component leads to an exponential suppression, governed precisely by the ``magnitude'' of this component (measured in terms of the metric $g_{\partial M}$).

We now turn to the context of axiom (T3x). Let $\Sigma$ be a hypersurface. Then $\Sigma$ defines an \emph{empty region} $\hat{\Sigma}$ with boundary $\partial\hat{\Sigma}=\overline{\Sigma}\cup\Sigma'$. Here, $\Sigma'$ denotes a second copy of $\Sigma$. The following Proposition shows that axiom (T3x) is satisfied.
\begin{prop}
\label{prop:ipampl}
We have $\cH_{\overline{\Sigma}}\tens\cH_{\Sigma'}\subseteq\cH_{\partial \hat{\Sigma}}^\ds$. Moreover, for $\psi,\psi'\in \cH_\Sigma$ we have,
\begin{equation}
 \rho_{\hat{\Sigma}}(\iota_\Sigma(\psi)\tens\psi')
=\langle\psi, \psi'\rangle_\Sigma .
\label{eq:ipampl}
\end{equation}
\end{prop}
\begin{proof}
We fix a base point $\eta\in A_{\Sigma}$. This yields the base point $(\eta,\eta)\in A_{\tilde{\hat{\Sigma}}}\subseteq A_{\partial \hat{\Sigma}}=A_{\overline{\Sigma}}\times A_{\Sigma'}$. The decompositions of the wave functions $\iota_\Sigma(\psi)\tens\psi'$, $\psi$ and $\psi'$ according to (\ref{eq:wfbase}) satisfy the equality
\begin{equation}
\chi^{(\eta,\eta)}(\phi,\phi)=\overline{\chi^\eta(\phi)}{\chi'}^{\eta}(\phi),
\end{equation}
where $\phi\in L_\Sigma$ with the obvious notation. This in turn implies
\begin{equation}
\int_{\hat{L}_{\tilde{\hat{\Sigma}}}} \chi^{(\eta,\eta)}(\tilde{\phi})\,\xd\nu_{\tilde{\hat{\Sigma}}}(\tilde{\phi})
=\int_{\hat{L}_\Sigma} \overline{\chi^\eta(\phi)}{\chi'}^{\eta}(\phi)\,\xd\nu_{\Sigma}(\phi) ,
\label{eq:erint}
\end{equation}
due to the equality of the measures $\nu_{\tilde{\hat{\Sigma}}}$ and $\nu_\Sigma$, when identifying $\tilde{\phi}\in L_{\tilde{\hat{\Sigma}}}\subseteq L_{\partial\hat{\Sigma}}=L_{\overline{\Sigma}}\times L_{\Sigma'}$ with $\phi\in L_\Sigma$ via $\tilde{\phi}=(\phi,\phi)$, see also Proposition~4.3 of \cite{Oe:holomorphic}.
But the right-hand side of (\ref{eq:erint}) is precisely the inner product (\ref{eq:hsip}) between $\psi$ and $\psi'$. On the other hand, the left-hand side of (\ref{eq:erint}) is precisely the amplitude (\ref{eq:defampl}) of $\iota_\Sigma(\psi)\tens\psi'$. To see this it remains to remark that $S_{\hat{\Sigma}}=0$ since $S_{\hat{\Sigma}}+S_{\hat{\Sigma}}=S_{\hat{\Sigma}}$ by axiom (C6). We obtain equation (\ref{eq:ipampl}). Note that this also implies $\iota_\Sigma(\psi)\tens\psi'\in\cH_{\partial \hat{\Sigma}}^\ds$ and hence $\cH_{\overline{\Sigma}}\tens\cH_{\Sigma'}\subseteq\cH_{\partial \hat{\Sigma}}^\ds$ as integrability on the right-hand side of (\ref{eq:erint}) implies integrability on the left-hand side.
\end{proof}

Let us return at this point to the question of the relative sign in equation (\ref{eq:relspact}) of Section~\ref{sec:classft}. As mentioned there, this has to do with the matching between geometric quantization of states spaces and the Feynman quantization of amplitudes. The issue becomes manifest precisely in the present Section. Indeed, everything concerning purely the construction of states and state spaces (Sections~\ref{sec:sspaces} and \ref{sec:cohstates}) is independent of the sign. Also, a large part of the present section would carry through (with modified formulas though) for the other choice of sign in equation (\ref{eq:relspact}). It is only via axioms (T3x) and (T5b) (to be considered in the next section) that the two quantization prescriptions are really fitted together. Indeed, Proposition~\ref{prop:ipampl} and Theorem~\ref{thm:gluing} are the instances which really require equation (\ref{eq:relspact}) in the given form.

\subsection{Gluing}
\label{sec:gluing}

We proceed in this section to demonstrate the validity of the gluing axiom (T5b), which we restate in a convenient form. Let $M$ be a region with its boundary decomposing as a disjoint union $\partial M=\Sigma_1\cup\Sigma\cup \overline{\Sigma'}$, where $\Sigma'$ is a copy of $\Sigma$. $M_1$ denotes the gluing of $M$ with itself along $\Sigma,\overline{\Sigma'}$ and we suppose that $M_1$ is an admissible region. We note $\partial M_1=\Sigma_1$. Fixing a base point $\eta\in A_\Sigma$ the axiom requires for all $\psi\in\cH_{\Sigma_1}^\ds$,
\begin{equation}
 \rho_{M_1}(\psi)\cdot c(M;\Sigma,\overline{\Sigma'})=\int_{\hat{L}_\Sigma}\rho_M(\psi\tens \hat{K}_{\eta+\xi}\tens\iota_\Sigma(\hat{K}_{\eta+\xi}))
\exp\left(\frac{1}{2}g_\Sigma(\xi,\xi)\right)\xd\nu_\Sigma(\xi) ,
\label{eq:glueeq}
\end{equation}
where $c(M;\Sigma,\overline{\Sigma'})$ is a non-zero complex number that only depends on the geometric data, called the \emph{anomaly factor}. As in \cite{Oe:holomorphic} we use here a completeness relation of coherent states to accomplish the gluing on the hypersurface $\Sigma$ rather than a sum over an orthonormal basis. This is equivalent due to the completeness relation (\ref{eq:cohcompl}) which follows from the corresponding completeness relation (4.1) in \cite{Oe:holomorphic}. The remark made above that the completeness relation (\ref{eq:cohcompl}) is independent of the choice of base point $\eta\in A_\Sigma$ applies here equally for the right-hand side of expression (\ref{eq:glueeq}).

In order to demonstrate the gluing axiom we will heavily rely on the proof for the linear theory given in \cite{Oe:holomorphic}. In particular, it will be convenient to recall the equation corresponding to (\ref{eq:glueeq}) in that context,
\begin{equation}
 \rho_{M_1}^{\mathrm{L}}(\psi)\cdot c(M;\Sigma,\overline{\Sigma'})=\int_{\hat{L}_\Sigma}\rho_M^{\mathrm{L}}(\psi\tens K_{\xi}\tens\iota_\Sigma(K_{\xi}))\, \xd\nu_\Sigma(\xi) .
\label{eq:gluelin}
\end{equation}
Here, we have marked the amplitude maps with a superscript $^{\mathrm{L}}$ to distinguish them from their counterparts in the affine context of the present paper. The anomaly factor given in \cite{Oe:holomorphic} will turn out to be the same here, so we do not need to distinguish it notationally. Indeed, recall from Theorem~4.5 of \cite{Oe:holomorphic},
\begin{equation}
c(M;\Sigma,\overline{\Sigma'})
=\int_{\hat{L}_\Sigma}\rho_M^{\mathrm{L}}(K_0\tens K_\xi\tens\iota_\Sigma(K_\xi))\,
 \xd\nu_\Sigma(\xi) .
\label{eq:aflin}
\end{equation}
We also recall that for this to make sense the integral on the right-hand side needs to exist. This was called the \emph{integrability condition} for the gluing data in \cite{Oe:holomorphic}.

In order to relate the amplitude maps in the present setting with those in the linear setting, we compare the definition of $\rho^{\mathrm{L}}$ from expression (4.4) in \cite{Oe:holomorphic} with equation (\ref{eq:defampl}). For a region $N$, a base point $\eta\in A_N$ and a state $\psi\in\cH_{\partial N}$ decomposed according to (\ref{eq:wfbase}) we obtain,
\begin{equation}
\rho_N(\psi)=\exp\left(\im S_N(\eta)\right)\rho_N^{\mathrm{L}}(\chi^\eta) .
\label{eq:relampl}
\end{equation}
In particular, for coherent states of the type exhibited in (\ref{eq:deccohwf}) we obtain
\begin{equation}
\rho_N(K^\eta_\xi)=\exp\left(\im S_N(\eta)\right) \rho_N^{\mathrm{L}}(K_\xi) .
\label{eq:relamplcoh}
\end{equation}
Thus, choosing a base point $\eta\in A_{M_1}$ and using equation (\ref{eq:relcohst}) we can rewrite the integrand of (\ref{eq:aflin}) in terms of the affine amplitude map and affine coherent states,
\begin{multline}
 \rho_M^{\mathrm{L}}(K_0\tens K_\xi\tens\iota_\Sigma(K_\xi))\\
= \exp\left(-\im S_M(\eta)+\frac{1}{2}g_\Sigma(\xi,\xi)\right)
  \rho_M(\hat{K}_\eta\tens\hat{K}_{\eta_0+\xi}\tens\iota_\Sigma(\hat{K}_{\eta_0+\xi})) .
\label{eq:relsampl}
\end{multline}
Here $\eta_0\in A_{\Sigma}$ denotes the element induced by $\eta\in A_{M_1}$ via axiom (C7), compare (\ref{eq:exsintbdy}). We thus state the integrability condition for gluing data in the present context as follows.

\begin{dfn}
\label{def:intcond}
We say that the gluing data satisfy the \emph{integrability condition} if
for some, hence any, $\eta\in A_{M_1}$, the extension of the function $\Phi:L_\Sigma\to\C$ defined by
\begin{equation}
 \Phi(\xi)\defeq \rho_M(\hat{K}_\eta\tens \hat{K}_{\eta_0+\xi}\tens\iota_\Sigma(\hat{K}_{\eta_0+\xi}))\exp\left(\frac{1}{2}g_\Sigma(\xi,\xi)\right)
\label{eq:intcond}
\end{equation}
to a function on $\hat{L}_\Sigma$ is $\nu_\Sigma$-integrable and its integral is different from zero.
\end{dfn}

The independence of integrability on the choice of the base point follows immediately from the equality (\ref{eq:relsampl}). With the additional assumption, apart from the already stated axioms for the classical data, that the integrability condition is satisfied for all admissible gluings it is now quite straightforward to show the validity of the gluing axiom (T5b) based on the corresponding Theorem~4.5 in \cite{Oe:holomorphic}.

\begin{thm}
\label{thm:gluing}
If the integrability condition is satisfied for all admissible gluings, then axiom (T5b) holds. Moreover, given a base point $\eta\in A_{M_1}$,
\begin{multline}
c(M;\Sigma,\overline{\Sigma'})
=\exp\left(-\im S_{M_1}(\eta)\right)\\
\int_{\hat{L}_\Sigma} \rho_M(\hat{K}_\eta\tens \hat{K}_{\eta_0+\xi}\tens\iota_\Sigma(\hat{K}_{\eta_0+\xi}))\exp\left(\frac{1}{2}g_\Sigma(\xi,\xi)\right) \xd\nu_\Sigma(\xi) ,
\label{eq:afaf}
\end{multline}
and this expression is independent of the choice of base point.
\end{thm}
\begin{proof}
The expression (\ref{eq:afaf}) just repeats the definition (\ref{eq:aflin}) in a form adapted to the present context, using (\ref{eq:relsampl}). This implies in particular its independence of the choice of base point. Note that $S_{M_1}(\eta)=S_M(\eta)$ in the light of axiom (C7). It remains to show that equation (\ref{eq:glueeq}) reduces to equation (\ref{eq:gluelin}). To this end, it is sufficient to demonstrate this for coherent states as their linear combinations are dense in $\cH_{\partial M_1}^\ds$. It will be slightly more convenient to the use the coherent states $K^\eta_\xi$ defined in (\ref{eq:deccohwf}) rather then the affine coherent states $\hat{K}_\zeta$ defined in (\ref{eq:defacoh}). However, in the light of equation (\ref{eq:relcohst}) these are equivalent for the present purposes. We use a base point $\zeta\in A_{M_1}$. Let $\phi\in L_{\partial {M_1}}$ be arbitrary. With (\ref{eq:relamplcoh}) and (\ref{eq:relcohst}) we obtain
\begin{align}
& \rho_{M_1}(K^\zeta_\phi)\cdot c(M;\Sigma,\overline{\Sigma'})\\
& =\exp\left(\im S_{M_1}(\zeta)\right) \rho_{M_1}^{\mathrm{L}}(K_\phi)\cdot c(M;\Sigma,\overline{\Sigma'})\\
& =\exp\left(\im S_{M_1}(\zeta)\right) \int_{\hat{L}_\Sigma}\rho_M^{\mathrm{L}}(K_\phi\tens K_\xi\tens\iota_\Sigma(K_\xi))\,
 \xd\nu_\Sigma(\xi)\\
& =\int_{\hat{L}_\Sigma}\rho_M(K_\phi^\zeta\tens K_\xi^{\zeta_0}\tens\iota_\Sigma(K_\xi^{\zeta_0}))\, \xd\nu_\Sigma(\xi)\\
& =\int_{\hat{L}_\Sigma}\rho_M(K_\phi^\zeta\tens \hat{K}_{\zeta_0+\xi}\tens\iota_\Sigma(\hat{K}_{\zeta_0+\xi}))\exp\left(\frac{1}{2}g_\Sigma(\xi,\xi)\right) \xd\nu_\Sigma(\xi)
\end{align}
It remains to observe that this is a special case of the right-hand side of (\ref{eq:glueeq}) with $\eta=\zeta_0$, where $\zeta_0\in A_\Sigma$ is induced from $\zeta$ by axiom (C7).
\end{proof}

%% file: prop.tex
\section{Further properties and extensions}
\label{sec:prop}

\subsection{Evolution picture}
\label{sec:evol}

We consider in this section the implications of the presented quantization scheme for the ``evolution'' of states between hypersurfaces. That is, we consider situations with regions where there is a one-to-one correspondence between classical solutions on one boundary component and those on another boundary component. This generalizes results of Section~4.5 of \cite{Oe:holomorphic}.

Let $M$ be a region such that its boundary decomposes as a disjoint union of two components $\partial M=\Sigma_1\cup\overline{\Sigma_2}$. Let the maps $a_1:A_{\tilde{M}}\to A_{\Sigma_1}$ and $a_2:A_{\tilde{M}}\to A_{\Sigma_2}$ be given by $a_{M;\Sigma_1,\overline{\Sigma_2}}$ with subsequent projection. Suppose that $a_1, a_2$ are invertible and such that the composition $T\defeq a_2\circ a_1^{-1}:A_{\Sigma_1}\to A_{\Sigma_2}$ is a homeomorphism. Informally speaking, we are considering the situation of a 1-1 correspondence between ``initial data'' on the hypersurfaces $\Sigma_1$ and $\Sigma_2$, mediated by the map $T$.

Due to the compatibility conditions in the classical axioms the corresponding ``linearized'' maps $r_1:L_{\tilde{M}}\to L_{\Sigma_1}$, $r_2:L_{\tilde{M}}\to L_{\Sigma_2}$ and $\tilde{T}\defeq r_2\circ r_1^{-1}:L_{\Sigma_1}\to L_{\Sigma_2}$ then have the same properties, in addition to being linear. We also have a correspondence between symplectic structures,
\begin{equation}
\omega_{\Sigma_2}(\tilde{T}\phi,\tilde{T}\phi')=\omega_{\Sigma_1}(\phi,\phi')\qquad\forall\phi,\phi'\in L_{\Sigma_1}
\label{eq:symevol}
\end{equation}
due to axiom (C5). However, we do not necessarily have
\begin{equation}
J_{\Sigma_2}\circ \tilde{T}=\tilde{T}\circ J_{\Sigma_1} .
\label{eq:jevol}
\end{equation}
But if (and only if) this is true, $\tilde{T}$ is unitary and we obtain a particularly ``nice'' evolution picture.

\begin{prop}
\label{prop:aevol}
There is a linear map $U:\cH_{\Sigma_1}\to\cH_{\Sigma_2}$ such that
\begin{equation}
 \rho_M(\psi_1\tens\iota_{\Sigma_2}(\psi_2))=\langle \psi_2,U\psi_1\rangle_{\Sigma_2}\qquad
 \forall \psi_1,\psi_2\in\cH_{\Sigma_2}.
\label{eq:ipamplevol}
\end{equation}
In particular, $U$ is given by
\begin{equation}
(U\psi)(\varphi)=\rho_M\left(\psi\tens \hat{K}_{\overline{\Sigma_2},\varphi}\right)
 \qquad\forall \psi\in\cH_{\Sigma_1},
  \forall\varphi\in A_{\Sigma_2}.
\label{eq:stateevol}
\end{equation}
Moreover, if $\tilde{T}$ is unitary then $U$ is unitary and we have
\begin{gather}
(U\psi)(\varphi)=\exp\left(\im S_M\left(a_2^{-1}(\varphi)\right)\right)\psi(T^{-1}\varphi)\qquad\forall \psi\in\cH_{\Sigma_1},
  \forall\varphi\in A_{\Sigma_2}. \label{eq:unitindev} \\
U \hat{K}_{\Sigma_1,\zeta}=\exp\left(\im S_M\left(a_1^{-1}(\zeta)\right)\right) \hat{K}_{\Sigma_2,T\zeta}\qquad \forall\zeta\in A_{\Sigma_1} .
\label{eq:evolcoh}
\end{gather}
\end{prop}
\begin{proof}
We rely here on the corresponding Proposition~4.6 in \cite{Oe:holomorphic}. We choose a base point $\eta\in A_M$ and denote its image under $a_M$ in $A_{\Sigma_1}\times A_{\overline{\Sigma_2}}$ by $(\eta_1,\eta_2)$. Taking (\ref{eq:stateevol}) as a definition and decomposing $\psi$ according to (\ref{eq:wfbase}) with base point $\eta_2$ we obtain,
\begin{align}
(U\psi)(\eta_2+\phi) & =\rho_M\left(\psi\tens\hat{K}_{\overline{\Sigma_2},\eta_2+\phi}\right) \label{eq:sevol1}\\
 & = \exp\left(\im S_M(\eta)+\im\theta_{\Sigma_2}(\eta_2,\phi)+\frac{\im}{2}[\phi,\phi]_{\Sigma_2}-\frac{1}{4}g_{\Sigma_2}(\phi,\phi)\right) \nonumber\\
 &\qquad \rho_M^{\mathrm{L}}\left(\chi^{\eta_1}\tens K^{\eta_2}_{\overline{\Sigma_2},\phi}\right) \label{eq:sevol2}\\
 & = \exp\left(\im S_M(\eta)\right) \alpha_{\Sigma_2}^{\eta_2}(\eta_2+\phi)
 \rho_M^{\mathrm{L}}\left(\chi^{\eta_1}\tens K^{\eta_2}_{\overline{\Sigma_2},\phi}\right)
\label{eq:sevol3}
\end{align}
Here we have used equations (\ref{eq:relcohst}) and (\ref{eq:relampl}) in the step from (\ref{eq:sevol1}) to (\ref{eq:sevol2}) and the definition (\ref{eq:defalpha}) in the step from (\ref{eq:sevol2}) to (\ref{eq:sevol3}). Let now $\chi_1^{\eta_1}$ and $\chi_2^{\eta_2}$ arise from the decompositions of $\psi_1$ and $\psi_2$ according to (\ref{eq:wfbase}) respectively. Then,
\begin{align}
\langle \psi_2,U\psi_1\rangle_{\Sigma_2}
& = \exp\left(\im S_M(\eta)\right) \int_{\hat{L}_{\Sigma_2}} \chi_2^{\eta_2}(\phi)\,
 \rho_M^{\mathrm{L}}\left(\chi_1^{\eta_1}\tens K^{\eta_2}_{\overline{\Sigma_2},\phi}\right)
 \,\xd\nu(\phi) \label{eq:ipevol1}\\
& = \exp\left(\im S_M(\eta)\right) \rho_M^{\mathrm{L}}\left(\chi_1^{\eta_1}\tens\iota_{\Sigma_2}^{\mathrm{L}}(\chi_2^{\eta_2})\right) \label{eq:ipevol2}\\
& = \rho_M(\psi_1\tens\iota_{\Sigma_2}(\psi_2)) .
 \label{eq:ipevol3}
\end{align}
Here we have used the first part of Proposition~4.6 of \cite{Oe:holomorphic} for the step from (\ref{eq:ipevol1}) to (\ref{eq:ipevol2}) and equation (\ref{eq:relampl}) for the step from (\ref{eq:ipevol2}) to (\ref{eq:ipevol3}).

We proceed to consider the special case that $\tilde{T}$ is unitary. Then, the next to last line of Proposition~4.6 of \cite{Oe:holomorphic} implies
\begin{equation}
 \rho_M^{\mathrm{L}}\left(\chi^{\eta_1}\tens K^{\eta_2}_{\overline{\Sigma_2},\phi}\right)
 =\chi^{\eta_1}(\tilde{T}^{-1}\phi) .
\end{equation}
Inserting this into (\ref{eq:sevol3}) yields,
\begin{align}
& (U\psi)(\eta_2+\phi)  = \exp\left(\im S_M(\eta)\right) \alpha_{\Sigma_2}^{\eta_2}(\eta_2+\phi)\, \chi^{\eta_1}(\tilde{T}^{-1}\phi) \label{eq:indevol1} \\
& = \exp\left(\im S_M(\eta)
 +\frac{\im}{2}\theta_{\Sigma_2}(\eta_2,\phi)
 +\frac{\im}{2}\theta_{\Sigma_2}(\eta_2+\phi,\phi)-\frac{1}{4}g_{\Sigma_2}(\phi,\phi)\right. \nonumber\\
&\qquad \left. -\frac{\im}{2}\theta_{\Sigma_1}(\eta_1,\tilde{T}^{-1}\phi)
 -\frac{\im}{2}\theta_{\Sigma_1}(\eta_1+\tilde{T}^{-1}\phi,\tilde{T}^{-1}\phi)
 +\frac{1}{4}g_{\Sigma_1}(\tilde{T}^{-1}\phi,\tilde{T}^{-1}\phi)\right) \nonumber\\
&\qquad  \alpha_{\Sigma_1}^{\eta_1}(\eta_1+\tilde{T}^{-1}\phi)\, \chi^{\eta_1}(\tilde{T}^{-1}\phi) \label{eq:indevol2} \\
& = \exp\left(\im S_M(\eta)-\frac{\im}{2}\theta_{\partial M}(\eta,\tilde{\phi})
 -\frac{\im}{2}\theta_{\partial M}(\eta+\tilde{\phi},\tilde{\phi})
 \right) \psi(\eta_1+\tilde{T}^{-1}\phi) \label{eq:indevol3} \\
& = \exp\left(\im S_M(\eta+\tilde{\phi})\right)  \psi(\eta_1+\tilde{T}^{-1}\phi) . \label{eq:indevol4}
\end{align}
We have used the definition (\ref{eq:defalpha}) of $\alpha$ in the step from (\ref{eq:indevol1}) to (\ref{eq:indevol2}), then used unitarity, collected the symplectic potential terms and used the decomposition of $\psi$ in the step to (\ref{eq:indevol3}). Here $\tilde{\phi}$ is a shorthand for the solution $(\tilde{T}^{-1}\phi,\phi)\in L_{\Sigma_1}\times L_{\overline{\Sigma_2}}=L_{\partial M}$. Then we have used equation (\ref{eq:actsympot}) to obtain (\ref{eq:indevol4}). The obtained equality can be conveniently rewritten as (\ref{eq:unitindev}).

Finally, to obtain (\ref{eq:evolcoh}) one uses the explicit expressions (\ref{eq:defacoh}) for the coherent state wave functions as well as (\ref{eq:actsympot}) again. We omit the straightforward details.
\end{proof}

Similar to the case of linear field theory, we can apply the above statement in a situation where there is a correspondence between solution spaces on any pair of admissible hypersurfaces to construct a quantization that implements unitary evolution between any such hypersurfaces by ``forwarding'' the complex structure with (\ref{eq:jevol}). See also the corresponding comments in \cite{Oe:holomorphic}.

Incidentally this type of setting provides a large class of examples for the axioms of Section~\ref{sec:classdata}. For concreteness consider a spacetime given by a globally hyperbolic manifold. On this spacetime consider a classical Lagrangian field theory with affine spaces of solutions of the Euler-Lagrange equations. Typically the latter would be (possibly inhomogeneous) hyperbolic partial differential equations. As usual suppose that any global solution is in one-to-one correspondence to initial data on any Cauchy hypersurface. If this theory can be quantized in a conventional way, then it can be made into an example of the axioms of Section~\ref{sec:classdata}.

To see this define a spacetime system as follows. Define the connected admissible hypersurfaces to be the the Cauchy hypersurfaces. Define the admissible hypersurfaces to be finite unions of non-intersecting connected admissible hypersurfaces. Define the regular connected admissible regions to be the submanifolds bounded by a pair of non-intersecting Cauchy hypersurfaces. Finally, define the regular admissible regions to be the finite unions of regular connected admissible regions which do not intersect in their interiors. The ingredients for the axioms of Section~\ref{sec:classdata} are defined as follows. Define the space $A_\Sigma$ associated to each Cauchy hypersurface $\Sigma$ to be the space of suitable initial data on that hypersurface with underlying vector space $L_\Sigma$. The Lagrangian setting yields the symplectic potential $\theta_\Sigma$ and its linearized version $[\cdot,\cdot]_\Sigma$. To each admissible region assign the space of suitable global solutions restricted to the region. Pick one particular hypersurface $\Sigma_0$ and define on $L_{\Sigma_0}$ a compatible complex structure $J_{\Sigma_0}$. (Here enters the assumption that the theory can be quantized in a ``conventional way''.) Forward this complex structure to any other Cauchy hypersurface through the correspondence of initial data.\footnote{This also clarifies what is meant by ``suitable'' initial data and ``suitable'' global solutions. In practice one would start with initial data satisfying certain restrictions (such as smoothness, decay properties etc.) and then completing the space of initial data with respect to the inner product induced by the complex structure.} It is not difficult to see that the assignment of data to admissible hypersurfaces and to admissible regions defined in this way satisfies precisely the axiomatic system of Section~\ref{sec:classdata}. Note also that in this setting the anomaly factor is always equal to one and the integrability condition is always satisfied.

\subsection{Vacuum}
\label{sec:vacuum}

Recall the vacuum axioms as presented in Section~2.3 of \cite{Oe:holomorphic} (and previously in Section~5.1 of \cite{Oe:2dqym}):

\begin{itemize}
\item[(V1)] For each hypersurface $\Sigma$ there is a distinguished state
  $\psi_{\Sigma,0}\in\cH_\Sigma$, called the \emph{vacuum state}.
\item[(V2)] The vacuum state is compatible with the involution. That is,
  for any hypersurface $\Sigma$,
  $\psi_{\bar{\Sigma},0}=\iota_\Sigma(\psi_{\Sigma,0})$.
\item[(V3)] The vacuum state is compatible with
  decompositions. Suppose the hypersurface
  $\Sigma$ decomposes into components
  $\Sigma_1\cup\dots\cup\Sigma_n$. Then
  $\psi_{\Sigma,0}=
 \tau_{\Sigma_1,\dots,\Sigma_n;\Sigma}(\psi_{\Sigma_1,0}\tens\cdots\tens\psi_{\Sigma_n,0})$.
\item[(V5)] The amplitude of the vacuum state is
  unity. That is, for any region $M$, $\rho_M(\psi_{\partial M,0})=1$.
\end{itemize}

In contrast to the linear theory treated in \cite{Oe:holomorphic}, the affine quantization scheme put forward here does not suggest a canonical vacuum. This is no surprise as the most natural vacuum in the linear case is given on the hypersurface $\Sigma$ by the coherent state $K_{\Sigma,0}$ associated to the special classical solution $0\in L_\Sigma$. It is precisely the absence of such a special point in $A_\Sigma$ that characterizes the present affine setting. On the other hand, it was pointed out in \cite{Oe:holomorphic} that any \emph{global solution} gives rise to a vacuum via coherent states. We proceed to recall this notion and show that it (almost) leads to a vacuum in the affine case as well.

To this end suppose that all regions and hypersurfaces of the spacetime system arise as submanifolds of a fixed manifold $B$ (possibly with additional structure) of dimension $d$. (But recall the related comments in Section~\ref{sec:geomax}.) Suppose now that there exists a solution $\varphi$ of the classical field equations in all of $B$. This induces a particular local solution in any region and on any hypersurface. The coherent states associated with these solutions then almost form a vacuum in the sense of the axioms. We formalize this as follows (compare Definition~4.7 of \cite{Oe:holomorphic}.)

\begin{dfn}
\label{dfn:gsol}
Let $\{\varphi_\Sigma\}$ be an assignment of an element $\varphi_\Sigma\in A_\Sigma$ to every hypersurface $\Sigma$. Then we call this assignment a \emph{global solution} iff it satisfies the following properties:
\begin{itemize}
\item[1.] Let $\Sigma$ be a hypersurface. Then, $\varphi_{\overline{\Sigma}}=\varphi_{\Sigma}$.
\item[2.] Suppose the hypersurface $\Sigma$ decomposes into a disjoint union of hypersurfaces $\Sigma=\Sigma_1\cup\cdots\cup\Sigma_n$. Then, $\varphi_\Sigma=(\varphi_{\Sigma_1},\dots,\varphi_{\Sigma_n})$.
\item[3.] Let $M$ be a region. Then, $\varphi_{\partial M}\in A_{\tilde{M}}$.
\end{itemize}
\end{dfn}

Setting $\psi_{\Sigma,0}\defeq \hat{K}_{\Sigma,\varphi_\Sigma}$ for every hypersurface $\Sigma$ satisfies (V1) by definition, (V2) due to property (\ref{eq:involcoh}), and (V3) due to property (\ref{eq:deccoh}). However, Proposition~\ref{prop:cohampl} yields in this situation for a region $M$,
\begin{equation}
 \rho_M(\psi_{\partial M,0})=\exp\left(\im S_M(\varphi_{\partial M})\right) .
\end{equation}
That is, axiom (V5) is generically not satisfied. However, the violation of axiom (V5) is ``mild'' in the sense that the amplitude has unit modulus, i.e., is ``merely'' a phase. One could take this as a motivation for weakening the axiom. After all, an overall phase factor does not contribute to the physics locally. We could thus postulate:
\begin{itemize}
\item[(V5')] The amplitude of the vacuum state is
  of unit modulus. That is, for any region $M$, $|\rho_M(\psi_{\partial M,0})|=1$.
\end{itemize}

However, there is also a way to satisfy the vacuum axioms in their present form by slightly modifying the quantization prescription. In fact, it is not really necessary to modify the quantization prescription itself, but merely one of its ingredients, the action. Indeed, it is sufficient to add, for each region $M$, a suitable constant to the action. Concretely, for a region $M$ we replace $S_M$ with
\begin{equation}
 \tilde{S}_M(\zeta)\defeq S_M(\zeta)-S_M(\varphi_{\partial M}) .
\label{eq:shiftact}
\end{equation}
It is then an easy exercise to verify that the axioms (C1)--(C7) are equally satisfied with the new action $\tilde{S}_M$. Also, the integrability condition, Definition~\ref{def:intcond}, remains unaffected. Classically, the actions $S_M$ and $\tilde{S}_M$ are of course completely equivalent. Quantum mechanically, the Feynman prescription (\ref{eq:pathint}) is also defined up to a normalization constant only and there is no reason a priori to fix the additive constant in the action one way or another.

\subsection{Reduction to the linear setting}
\label{sec:redlin}

The relationship of affine field theory and its quantization as presented here to linear field theory and its quantization as presented in \cite{Oe:holomorphic} has been a recurring theme in much of the discussion so far. Moreover, much of the results pertaining to the affine setting have been shown by recurrence to results in the linear setting. Nevertheless, let us spell out here in a more precise manner in which sense the affine setting can be reduced to the linear one. 

On the classical side, it is clear that to make the spaces of solutions associated to regions or hypersurfaces into vector spaces we merely need to choose base points. Of course, the difficulty lies in the fact that these choices have to be consistent under decompositions, gluings etc. In the special context of a global background, we have, however, just seen (in Section~\ref{sec:vacuum}) how to formalize this consistently through the notion of a global solution (Definition~\ref{dfn:gsol}).

Given the existence of a global solution, we may bring the present quantization scheme in one-to-one correspondence with that presented in \cite{Oe:holomorphic} for linear field theory. As a first step, we adapt the action to the linear case by performing the substitution indicated in equation (\ref{eq:shiftact}). This amounts to declaring that the action applied to the solution $\varphi_{\partial M}$, which is to represent the $0$, vanishes. For wave functions on a hypersurface $\Sigma$ the correspondence between the affine ones $\psi$ and the linear ones $\chi^{\varphi_\Sigma}$ is given by equation (\ref{eq:wfbase}), where the base point is $\varphi_\Sigma$. The coherent states $K_\xi$ of the linear setting are then recovered as $K_\xi^{\varphi_\Sigma}$ defined by formula (\ref{eq:deccohwf}). For amplitudes we obtain equality for corresponding states due to formula (\ref{eq:relampl}), where the base point is now $\varphi_{\partial M}$.

Of course the situation considered here is very special. Not having a global solution, or not even having a global background means that there is no such coherent choice of base points and hence no reduction. Reduction is also precluded if we are interested not in a single affine theory, but in a family of such, perhaps even separately for each region (see Section~\ref{sec:linsrc} for such a setting). So the affine theory is really more general and, even in a reducible situation, often more natural than the linear one. Moreover, the affine coherent states really differ from the linear ones, even given the choice of a base point. Their principal advantage over the linear ones is lack of reference to such a base point, see (\ref{eq:ipacoh}).

Finally let us stress a point that was left unclear in \cite{Oe:holomorphic}. There, the amplitude map (equation (4.4) in that paper) was postulated without much further justification. In the present paper we have clarified how it is motivated, as a special case of the definition (\ref{eq:defampl}), through the Feynman path integral formula (Section~\ref{sec:feynquant}). In doing so, we have also fixed the relative sign in equation (\ref{eq:relspact}). This sign was implicitly left open in \cite{Oe:holomorphic}, even though in the applications of Section~5 of that paper, it was indeed taken to be negative in order to achieve agreement with known quantizations of Klein-Gordon theory in certain geometries.

\subsection{Observables}
\label{sec:observables}

The Berezin-Toeplitz quantization of observables put forward for the linear setting in \cite{Oe:obsgbf} can be straightforwardly generalized to the affine setting. We sketch this in the present section.

 We model a classical observable $F$ on a spacetime region $M$ as a map $A_{\tilde{M}}\to\C$ (or $A_{\tilde{M}}\to\R$) and define the associated quantized observable map, given a base point $\eta\in A_M$, via
\begin{equation}
 \rho_M^{\ano{F}}(\psi)\defeq\exp\left(\im S_M(\eta)\right)\int_{\hat{L}_{\tilde{M}}} F(\eta+\phi)\chi^\eta(\phi)\,\xd\nu_{\tilde{M}}(\phi) .
\label{eq:defobs}
\end{equation}
Here $\chi^\eta$ comes from the decomposition of $\psi$ according to equation (\ref{eq:wfbase}). To clarify this definition and for later use we recall a few technical details from \cite{Oe:obsgbf}, adapting them to the present setting.
For a map $F:A_{\tilde{M}}\to\C$ and an element $\zeta\in A_{\tilde{M}}$ we denote by $F^\zeta:L_{\tilde{M}}\to\C$ the translated map $\phi\mapsto F(\zeta+\phi)$. We say that $F:A_{\tilde{M}}\to\C$ is \emph{analytic} iff for each pair $(\varphi,\xi)\in A_{\tilde{M}}\times L_{\tilde{M}}$ the map $z\mapsto F(\varphi+z\xi)$ is real analytic. We denote the induced extension $A_{\tilde{M}}^\C\to\C$ of $F$ also by $F$, where $A_{\tilde{M}}^\C=A_{\tilde{M}}\oplus \im L_{\tilde{M}}$ is the ``complexification'' of $A_{\tilde{M}}$. We say that $F:A_{\tilde{M}}\to\C$ is analytic and \emph{sufficiently integrable} iff for any $\zeta\in A_{\tilde{M}}^\C$ (the extension of) the map $F^\zeta$ is integrable in $(\hat{L}_{\tilde{M}},\nu_{\tilde{M}})$. In order for (\ref{eq:defobs}) to make sense we shall require $F$ to be analytic and sufficiently integrable. As we will see below, this guarantees the existence of (\ref{eq:defobs}) at least for coherent states.

A straightforward modification of the proof of Lemma~\ref{lem:amplibp}, which we leave to the reader, yields then the following,
\begin{lem}
The above definition of $\rho_M^{\ano{F}}(\psi)$ is independent of the choice of base point.
\end{lem}
Recall from Proposition~4.1 of \cite{Oe:holomorphic} that the linear analogue of the quantization formula (\ref{eq:defobs}) has the \emph{coherent factorization property}. That is, the quantized observable evaluated on a coherent state factorizes into an ordinary amplitude for the coherent state and a ``vacuum expectation value''. A very similar statement is true in the present affine setting, which we shall denote in the same way. To state it we first define for $\tau\in A_{\tilde{M}}^\C$ the following quantity,
\begin{equation}
\sigma_M^{\ano{F^{\tau}}}\defeq \int_{\hat{L}_{\tilde{M}}} F(\tau+\phi)\,\xd\nu_{\tilde{M}}(\phi) .
\end{equation}

\begin{prop}[Coherent Factorization Property]
Let $F:A_M\to\C$ be analytic and sufficiently integrable. Then, for any $\zeta\in A_{\partial M}$ we have
\begin{equation}
 \rho_M^{\ano{F}}(\hat{K}_\zeta)=\rho_M(\hat{K}_\zeta)\, \sigma_M^{\ano{F^{\hat{\zeta}}}},
\label{eq:cohfact}
\end{equation}
where $\hat{\zeta}\in A_{\tilde{M}}^{\C}$ is given by $\hat{\zeta}=\zeta^{\mathrm{R}}-\im\zeta^{\mathrm{I}}$.
\end{prop}
\begin{proof}
We reduce (\ref{eq:cohfact}) to the respective statement in the linear setting, Proposition~4.1 of \cite{Oe:obsgbf}. Fix a base point $\eta\in A_M$ and set $\xi\defeq\zeta-\eta$. Then,
\begin{align}
 \rho_M^{\ano{F}}(K^\eta_\xi) & =\exp\left(\im S_M(\eta)\right)
 \rho_M^{\mathrm{L}\,\ano{F^\eta}}(K_\xi)
 \label{eq:obsfac1}\\
& =\exp\left(\im S_M(\eta)\right)\rho_M^{\mathrm{L}}(K_\xi)\rho_M^{\mathrm{L}\,\ano{F^{\eta+\hat{\xi}}}}(K_0)\label{eq:obsfac2}\\
& =\rho_M(K^\eta_\xi)\rho_M^{\mathrm{L}\,\ano{F^{\hat{\zeta}}}}(K_0)\label{eq:obsfac3}\\
& =\rho_M(K^\eta_\xi)\sigma_M^{\ano{F^{\hat{\zeta}}}}\label{eq:obsfac4} .
\end{align}
Here we use the notation $\rho_M^{\mathrm{L}\,\ano{G}}$ to denote the quantization of the classical observable $G:L_{\tilde M}\to\C$ in the linear setting of \cite{Oe:obsgbf}. Thus, (\ref{eq:obsfac1}) is the analogue of equation (\ref{eq:relamplcoh}) for observables. The step from (\ref{eq:obsfac1}) to (\ref{eq:obsfac2}) is the application of Proposition~4.1 of \cite{Oe:obsgbf}. Here, $\hat{\xi}\defeq\xi^{\mathrm{R}}-\im \xi^{\mathrm{I}}$. The step from (\ref{eq:obsfac2}) to (\ref{eq:obsfac3}) is provided by an application of equation (\ref{eq:relamplcoh}) and the recognition that $\hat{\zeta}=\eta+\hat{\xi}$. The step from (\ref{eq:obsfac3}) to (\ref{eq:obsfac4}) arises from the recognition that $\sigma_M^{\ano{F^{\hat{\zeta}}}}$ coincides with $\rho_M^{\mathrm{L}\,\ano{F^{\hat{\zeta}}}}(K_0)$. To obtain (\ref{eq:cohfact}) we recall that $\hat{K}_\zeta$ is simply a multiple of $K^\eta_\xi$, see (\ref{eq:relcohst}).
\end{proof}

%% file: linsrc.tex
\section{``Asymptotically'' linear field theory}
\label{sec:linsrc}

In this section we consider a particular application of the quantization scheme put forward in the present paper. Suppose we are in a setting of classical Lagrangian field theory as outlined in Section~\ref{sec:classft}. For a fixed spacetime region $M$ we are given an action $S_M^\mu:K_M\to\R$ as a sum of two terms ($K_M$ denotes the vector space of field configurations in $M$),
\begin{equation}
 S_M^\mu(\phi)=S_M(\phi)+C_M^\mu(\phi) .
\label{eq:actsum}
\end{equation}
The first term, $S_M$, is quadratic in $\phi$, while the second term, $C_M^\mu$, is linear in $\phi$. Moreover, we shall assume that $C_M^\mu$ vanishes near the boundary of $M$.

The action $S_M^\mu$ thus defines an affine field theory in the sense of Section~\ref{sec:affine} and we shall use the notation introduced there. Thus we denote by $A_M$ and $A_{\partial M}$ the spaces of solutions in $M$ and near $\partial M$ respectively, and by $L_M$ and $L_{\partial M}$ their linear counter parts etc. On the other hand, $A_M$ and $A_{\partial M}$ may be viewed as subsets of $K_M$ and $K_{\partial M}$ (the latter being the vector space of field configurations near $\partial M$) and the linear spaces $L_M$ and $L_{\partial M}$ may be viewed as linear subspaces of $K_M$ and $K_{\partial M}$. Moreover, since $C_M^\mu$ vanishes near the boundary, the theory is linear there, meaning that $A_{\partial M}$ and $L_{\partial M}$ are really identified as subsets of $K_{\partial M}$. We shall freely use this identification in the following. Note that this identification is in general \emph{not} of the type as that obtained in Section~\ref{sec:redlin} via a coherent choice of base points. We may also view $S_M$ as defining a field theory in its own right, which is linear and whose spaces of solutions are precisely $L_M$ on $M$ and $L_{\partial M}$ near $\partial M$ (viewed again as subspaces of $K_M$ and $K_{\partial M}$ respectively).

We make the usual assumptions concerning the non-degeneracy of the symplectic form on the boundary and $L_M$ giving rise to a Lagrangian subspace of $L_{\partial M}$. For simplicity of notation we shall not explicitly distinguish between $A_{\tilde{M}}$ and $A_M$ or between $L_{\tilde{M}}$ and $L_M$.
Leaving out as usual the explicit mention of maps $a_M$ and $l_M$, equation (\ref{eq:propaact}) may be rewritten as,
\begin{equation}
 S_M^\mu(\eta+\xi)=S_M^\mu(\eta)-\frac{1}{2}[\xi,\xi]_{\partial M}-[\eta,\xi]_{\partial M} \qquad \forall \eta\in A_M, \forall \xi\in L_M .
\label{eq:xaact}
\end{equation}
On the other hand, applying equation (\ref{eq:propaact}) to $S_M$ and using the fact that it is quadratic yields,
\begin{equation}
 S_M(\xi)=-\frac{1}{2}[\xi,\xi]_{\partial M} \qquad\forall\xi\in L_M .
\label{eq:xlact}
\end{equation}

The affine theory determined by $S_M^\mu$ and the linear theory determined by $S_M$ define two different decompositions of the boundary solution space $A_{\partial M}=L_{\partial M}$ according to Lemmas~\ref{lem:decbdyas} and \ref{lem:decis} respectively. We shall write the decomposition of $\phi\in L_{\partial M}$ according to the affine theory as $\phi=\phi^{\bar{\mathrm{R}}}+J_{\partial M}\phi^{\bar{\mathrm{I}}}$ and according to the linear theory simply as $\phi=\phi^{\mathrm{R}}+J_{\partial M}\phi^{\mathrm{I}}$.\footnote{By slight abuse of notation we write $J_{\partial M}\eta^{\mathrm{I}}$ to also denote the solution $\eta-\eta^{\mathrm{R}}\in A_M$.}
In order to compare the two decompositions, choose an arbitrary element $\eta\in A_M$. Given $\phi\in L_{\partial M}$ set $\xi\defeq \phi-\eta$. Then we have on the one hand,
\begin{equation}
 \phi^{\bar{\mathrm{R}}}=\eta+\xi^{\mathrm{R}}\quad\text{and}\quad
 \phi^{\bar{\mathrm{I}}}=\xi^{\mathrm{I}} .
\end{equation}
On the other hand we have,
\begin{equation}
 \phi^{\mathrm{R}}=\eta^{\mathrm{R}}+\xi^{\mathrm{R}}\quad\text{and}\quad
 \phi^{\mathrm{I}}=\eta^{\mathrm{I}}+\xi^{\mathrm{I}} .
\end{equation}
This implies,
\begin{equation}
 \phi^{\bar{\mathrm{R}}}=\phi^{\mathrm{R}}+J_{\partial M}\eta^{\mathrm{I}}\quad\text{and}\quad
 \phi^{\bar{\mathrm{I}}}=\phi^{\mathrm{I}}-\eta^{\mathrm{I}} .
\label{eq:transdec}
\end{equation}
The apparent dependence in (\ref{eq:transdec}) on $\eta$ might seem disturbing until we realize that $\eta^{\mathrm{I}}$ does not actually depend on $\eta$. Indeed, $\eta^{\mathrm{I}}$ is the unique element of $L_M$ such that $J_{\partial M}\eta^{\mathrm{I}}\in A_M$. (This characterization assumes $A_M\neq L_M$. If this is not the case, then $\eta^{\mathrm{I}}=0$.)

Before proceeding we shall derive two additional identities under the assumption that there is an element $\eta_0\in A_M$ such that $[\xi,\eta_0]_{\partial M}=0$ for all $\xi\in L_M$. In other words, we assume that there is a solution $\eta_0$ of the affine theory that vanishes on the boundary of $M$, compare (\ref{eq:sympot}). From equations (\ref{eq:actsum}), (\ref{eq:xaact}), and (\ref{eq:xlact}) we infer,
\begin{equation}
S_M(\eta+\xi)=S_M(\xi)+S_M(\eta)-C_M^\mu(\xi)-[\eta,\xi]_{\partial M} \qquad \forall \eta\in A_M, \forall \xi\in L_M .
\end{equation}
On the other hand, viewing $\eta$ as a perturbation of the solution $\xi$ of the theory defined by $S_M$, the variational principle together with the fact that $S_M$ is quadratic implies,
\begin{equation}
S_M(\eta+\xi)=S_M(\xi)+S_M(\eta)+X_M(\xi,\eta) \qquad \forall \eta\in K_M, \forall \xi\in L_M ,
\end{equation}
where $X_M(\xi,\eta)$ is linear both in $\xi$ and the perturbation $\eta$ and moreover has the property that it vanishes if $\eta$ vanishes on the boundary $\partial M$. Thus,
\begin{equation}
X_M(\xi,\eta)=-C_M^\mu(\xi)-[\eta,\xi]_{\partial M} \qquad \forall \eta\in A_M, \forall \xi\in L_M .
\end{equation}
We rewrite this as,
\begin{equation}
X_M(\xi,\eta)=-C_M^\mu(\xi)+2\omega_{\partial M}(\xi,\eta)-[\xi,\eta]_{\partial M} \qquad \forall \eta\in A_M, \forall \xi\in L_M .
\label{eq:bilinpert}
\end{equation}
We note that neither the first nor the second term on the right hand side depend on $\eta$. To see this for the second term note that
\begin{equation}
\omega_{\partial M}(\xi,\eta')-\omega_{\partial M}(\xi,\eta)=\omega_{\partial M}(\xi,\eta'-\eta)=0
\end{equation}
for $\eta',\eta\in A_M$ since $\eta'-\eta\in L_M$ and $L_M$ is Lagrangian in $L_{\partial M}$. On the other hand, the third term in (\ref{eq:bilinpert}) has the required properties and vanishes if $\eta$ vanishes on the boundary $\partial M$, compare the explicit definition (\ref{eq:sympot}). Given the existence of $\eta_0\in A_M$ as described above we conclude that the sum of the first two terms on the right hand side of (\ref{eq:bilinpert}) must vanish, i.e.,
\begin{equation}
 C_M^\mu(\xi)=2\omega_{\partial M}(\xi,\eta) \qquad \forall\xi\in L_M .
\label{eq:srcsympl}
\end{equation}
As we have just previously shown, even though an element $\eta\in A_M$ appears on the right hand side, the expression is independent of the choice of this element.

Now let $\eta\in A_M$ and consider for $\lambda\in\R$,
\begin{equation}
S_M^\mu(\eta+\lambda\eta)=S_M^\mu(\eta)+\lambda(2 S_M(\eta)+C_M^\mu(\eta))+ \lambda^2 S_M(\eta),
\end{equation}
where we have used (\ref{eq:actsum}) and the fact that $S_M$ is quadratic while $C_M^\mu$ is linear. Viewing $\lambda\eta$ as a perturbation of the solution $\eta$, the term linear in $\lambda$ on the right hand side must vanish if $\eta$ vanishes on the boundary. That this can happen is ensured by the existence of $\eta_0$, yielding,
\begin{equation}
S_M(\eta_0)=-\frac{1}{2} C_M^\mu(\eta_0) .
\end{equation}
Using (\ref{eq:actsum}), (\ref{eq:xaact}) and (\ref{eq:xlact}) we can deduce from this for all $\eta\in A_M$,
\begin{equation}
S_M^\mu(\eta)=\frac{1}{2} C_M^\mu(\eta)-\frac{1}{2}[\eta,\eta]_{\partial M} .
\label{eq:aactsol}
\end{equation}

\subsection{Factorization of the amplitude}
\label{sec:factampl}

We proceed to evaluate the amplitude of a coherent state on the boundary for the affine theory determined by $S_M^\mu$ with the goal to compare it to the amplitude of the linear theory determined by $S_M$. We shall denote the amplitude of the former theory by $\rho_M^\mu$, while denoting the amplitude of the latter theory by $\rho_M$. We consider natural coherent states $K_\xi$ of the linear theory with $\xi\in L_{\partial M}$ on the boundary, which from the affine point of view can be identified as $K_\xi= K_\xi^0$, compare equation (\ref{eq:deccohwf}). Equation (\ref{eq:cohampl}) of Proposition~\ref{prop:cohampl} together with equation (\ref{eq:relcohst}) yields,
\begin{multline}
 \rho_M^\mu(K_\xi) =
\exp\left(\im S_M^{\mu}\left(\xi^{\bar{\mathrm{R}}}\right)-\im\,[\xi^{\bar{\mathrm{R}}},J_{\partial M}\xi^{\bar{\mathrm{I}}}]_{\partial M}-\frac{\im}{2}[J_{\partial M}\xi^{\bar{\mathrm{I}}},J_{\partial M}\xi^{\bar{\mathrm{I}}}]_{\partial M}\right. \\
\left.
-\frac{1}{2} g_{\partial M}\left(\xi^{\bar{\mathrm{I}}},\xi^{\bar{\mathrm{I}}}\right)+\frac{\im}{2}[\xi,\xi]_{\partial M}+\frac{1}{4}g_{\partial M}(\xi,\xi)\right) .
\end{multline}
Inserting the substitutions (\ref{eq:transdec}) and using (\ref{eq:xaact}) as well as standard identities leads to,
\begin{multline}
 \rho_M^\mu(K_\xi) =
\exp\left(\frac{1}{4} g_{\partial M}\left(\xi^{\mathrm{R}},\xi^{\mathrm{R}}\right)-\frac{1}{4} g_{\partial M}\left(\xi^{\mathrm{I}},\xi^{\mathrm{I}}\right)
 -\frac{\im}{2}g_{\partial M}\left(\xi^{\mathrm{R}},\xi^{\mathrm{I}}\right)\right.\\
+2\im\omega_{\partial M}(\xi^{\mathrm{R}},J_{\partial M}\eta^{\mathrm{I}})
+2\omega_{\partial M}(\xi^{\mathrm{I}},J_{\partial M}\eta^{\mathrm{I}})\\
\left. +\im S_M^{\mu}\left(J_{\partial M}\eta^{\mathrm{I}}\right)
+\frac{\im}{2} [J_{\partial M}\eta^{\mathrm{I}},J_{\partial M}\eta^{\mathrm{I}}]_{\partial M}
-\omega_{\partial M}(\eta^{\mathrm{I}},J_{\partial M}\eta^{\mathrm{I}})\right) .
\end{multline}
Examining this expression, we find that the terms in the first line are precisely those arising as the amplitude of the state $K_\xi$ in the linear theory given by $S_M$. Also, the terms in the first two lines vanish if we set $\xi=0$, i.e, if we evaluate on the vacuum state $K_0$. Thus, the third line, which is independent of $\xi$, represents the amplitude of the affine theory for the vacuum state. The second line on the other hand can be rewritten in the light of the identity (\ref{eq:srcsympl}), noticing that $J_{\partial M}\eta^{\mathrm{I}}\in A_M$. We arrive at the following identity,
\begin{equation}
 \rho_M^\mu(K_\xi) =\rho_M(K_\xi)
\exp\left(\im C_M^\mu(\xi^{\mathrm{R}}-\im \xi^{\mathrm{I}})\right)
\rho_M^\mu(K_0) ,
\label{eq:factampls}
\end{equation}
where we have extended $C_M^\mu$ to the complexified solution space $L_M^\C$ (or configuration space $K_M^\C$). We also find, using (\ref{eq:aactsol}),
\begin{equation}
\rho_M^\mu(K_0)=\exp\left(\frac{\im}{2} C_M^\mu(J_{\partial M}\eta^{\mathrm{I}})
-\frac{1}{2}C_M^\mu(\eta^{\mathrm{I}})\right) .
\label{eq:avacev}
\end{equation}
We note that the real part of the argument of the exponential is negative definite, since
\begin{equation}
-\frac{1}{2}C_M^\mu(\eta^{\mathrm{I}})=-\frac{1}{2} g_{\partial M}(\eta^{\mathrm{I}},\eta^{\mathrm{I}})
\end{equation}
meaning in physical terms that we obtain an exponential damping of the amplitude induced by the ``magnitude'' of the ``failure'' of $\eta$ to be a solution of the linear theory.

The remarkable factorization identity (\ref{eq:factampls}) is not quite unexpected. Indeed, there is another, conceptually distinct way, to arrive at the same identity, which we shall only sketch here. Recall that we have implemented the amplitude as a version of the Feynman path integral (\ref{eq:pathint}). In light of the decomposition (\ref{eq:actsum}) of the action we may view this as a path integral for the action of the linear theory, while the contribution from the term $C_M^\mu$ is viewed as the insertion of the extra factor
\begin{equation}
 F(\zeta)\defeq\exp\left(\im C_M^\mu(\zeta)\right)
\label{eq:expc}
\end{equation}
into the integral. Rather than interpret this as modifying the action $S_M$ we can interpret this as giving rise to quantization of the classical observable $F$
along the lines of \cite{Oe:obsgbf}. The corresponding \emph{quantum observable map} $\rho_M^F:\cH_{\partial M}\to\C$ is then essentially the same object as the amplitude map $\rho_M^\mu:\cH_{\partial M}\to\C$, but with an a priori different interpretation. The coherent factorization property for Feynman quantization in the linear setting \cite{Oe:obsgbf} then yields,
\begin{equation}
 \rho_M^F(K_\xi)=\rho_M(K_\xi) \rho_M^{F^{\hat{\xi}}}(K_0) ,
\label{eq:cohfactf}
\end{equation}
where $\hat{\xi}=\xi^{\mathrm{R}}-\im \xi^{\mathrm{I}}$ and $F^{\hat{\xi}}(\phi)=F(\hat{\xi}+\phi)$. We note that $F(\hat{\xi}+\phi)=F(\hat{\xi}) F(\phi)$ due to the explicit form (\ref{eq:expc}) of $F$. This implies in turn that we can decompose the second factor in (\ref{eq:cohfactf}) to arrive at,
\begin{equation}
 \rho_M^F(K_\xi)=\rho_M(K_\xi) F(\hat{\xi})\rho_M^{F}(K_0) ,
\end{equation}
which corresponds precisely to (\ref{eq:factampls}). Let us emphasize, however, that we have not given here a rigorous and general definition of a Feynman quantization of observables. In particular, in contrast to the treatment via the affine theory, we have not discussed the actual value of the quantity $\rho_M^{F}(K_0)$. In turns out though, that for the simple observables considered here this quantity does coincide with $\rho_M^\mu(K_0)$ found above and given in (\ref{eq:avacev}). We shall consider these issues elsewhere in more depth.

\subsection{Linear field theory with source}
\label{sec:musrc}

We proceed to remark on a use of the identity (\ref{eq:factampls}) that justifies the title of the present section. The notation involving $\mu$ is meant to suggest that $\mu$ is a source and the term $C_M^\mu$ takes a form as follows,
\begin{equation}
 C_M^\mu(\phi)=\int_M \mu(x)\cdot\phi(x)\,\xd^d x .
\end{equation}
Here $\cdot$ might be a scalar multiplication or it might involve a summation over internal indices. $\xd^d x$ is some spacetime volume form, which alternatively could be absorbed into $\mu$. Both $\eta^{\mathrm{I}}$ and $C_M^\mu$ become linear in $\mu$. Thus, the argument of the exponential in the middle term on the right hand side of (\ref{eq:factampls}) is linear in $\mu$, while the argument of the exponential in the ``vacuum expectation value'' (\ref{eq:avacev}) is quadratic in $\mu$.

Of particular interest are theories where the spaces $L_M$ are spaces of solutions of homogeneous partial differential equations on Lorentzian (or Riemannian) manifolds. Introducing a source $\mu$ as above then makes the spaces $A_M$ spaces of solutions of the corresponding inhomogeneous equations with precisely this source.
It was in such a context that a special case of the factorization identity (\ref{eq:factampls}) was first encountered \cite{CoOe:spsmatrix,CoOe:smatrixgbf}. There, a Schrödinger-Feynman quantization of Klein-Gordon theory in Minkowski space was considered for two types of regions: On the one hand a time-interval extended over all of space as in conventional transition amplitudes (allowing comparison with well known results) and on the other hand a ball of fixed radius extended over all of time. In both cases a formula with precisely the structure of (\ref{eq:factampls}) was found and the ``vacuum expectation value'' $\rho_M^\mu(K_0)$ was more specifically found to take the form
\begin{equation}
\rho_M^\mu(K_0)=\exp\left(\frac{\im}{2}\int_M\mu(x)G_F(x,x')\mu(x')\,\xd^4\, x\xd^4 x'\right) ,
\label{eq:vevkg}
\end{equation}
where $G_F$ is the Feynman propagator. Corresponding results were subsequently obtained in de~Sitter spacetime \cite{Col:desitterletter,Col:desitterpaper} and confirmed in a more general framework for scalar quantum field theory in curved spacetime \cite{CoDo:smatrixcsp}.
A more detailed discussion of the relationship between the present results and those of \cite{CoOe:spsmatrix,CoOe:smatrixgbf,Col:desitterletter,Col:desitterpaper,CoDo:smatrixcsp} requires to explore precisely the relationship between the holomorphic and Schrödinger representation among other things. We shall do this elsewhere. For the moment let us merely mention that (\ref{eq:vevkg}) coincides exactly with (\ref{eq:avacev}) for the case of the time-interval region and the standard complex structure of Klein-Gordon theory, while the case of the other type of region is more complicated.

To stress the significance of the amplitude function $\rho_M^\mu$, we recall from \cite{CoOe:spsmatrix,CoOe:smatrixgbf} that it can be used as a generating functional for the amplitude of an interacting field theory with interaction of the form
\begin{equation}
 S_M^{\mathrm{int}}(\phi)=\int_M V(x,\phi(x))\,\xd^d x .
\end{equation}
We assume here that the potential $V$ vanishes near the boundary of $M$. From a Feynman path integral point of view the amplitude $\rho_M^V$ of the theory determined by $S_M+S_M^{\mathrm{int}}$ is then formally given by
\begin{equation}
\rho_{M}^V(\psi) =  \exp\left(\im\int
V\left(x,-\im\frac{\delta}{\delta \mu(x)}\right)\xd^d x\right)
\rho_M^\mu(\psi)\bigg|_{\mu=0} .
\label{eq:pertampl}
\end{equation}
Moreover, as shown in \cite{CoOe:spsmatrix,CoOe:smatrixgbf}, taking a limit of regions we can extract the perturbative (non-renormalized) S-matrix from this expression.

\subsection{Evolution Picture}
\label{sec:evolal}

We turn to the special type of geometry considered in Section~\ref{sec:evol}, which permits an interpretation in terms of ``evolution'' between hypersurfaces. Thus, we suppose the boundary of the region $M$ decomposes into a disjoint union $\partial M=\Sigma_1\cup\overline{\Sigma_2}$. Moreover, we shall suppose that the linear theory determined by $S_M$ admits a unitary map $\tilde{T}:L_{\Sigma_1}\to L_{\Sigma_2}$ given by $\tilde{T}=r_2\circ r_1^{-1}$ with $r_1:L_{\tilde{M}}\to L_{\Sigma_1}$, $r_2:L_{\tilde{M}}\to L_{\Sigma_2}$ the canonical projections. Recall that this means ``conservation'' both of the symplectic structure (\ref{eq:symevol}) and of the complex structure (\ref{eq:jevol}). Under these circumstances it turns out that ``evolution'' in the affine theory determined by $S_M^\mu$ is given by a homeomorphism $T:A_{\Sigma_1}\to A_{\Sigma_2}$ whose linearization is $\tilde{T}$. While this is given by a composition $T=a_2\circ a_1^{-1}$ (compare Section~\ref{sec:evol}), a simple way to obtain $T$ is through the relation between the decompositions of the spaces $A_M$ and $L_M$ according to (\ref{eq:transdec}).

In order to simplify notation we shall identify the spaces $L_{\Sigma_1}$, $L_{\Sigma_2}$ and $L_{\tilde{M}}$ via the isomorphisms $r_1$, $r_2$ and $\tilde{T}$, denoting them as $L_\Sigma$. Similarly, we write $J$ for $J_{\Sigma_1}$ or $J_{\Sigma_2}$. For $\phi\in L_\Sigma$ we are interested in $T\phi\in L_\Sigma$ such that $(\phi,T\phi)\in A_{\tilde{M}}$. This is uniquely determined by the condition $(\phi,T\phi)^{\bar{\mathrm{I}}}=0$ in terms of the decomposition of Lemma~\ref{lem:decbdyas}. What is simple in the present setting is the decomposition in terms of Lemma~\ref{lem:decis},
\begin{equation}
 (\phi,T\phi)^{\mathrm{R}}=\frac{1}{2}(\phi+T\phi)\quad\text{and}\quad
 (\phi,T\phi)^{\mathrm{I}}=-\frac{1}{2}J(\phi-T\phi) .
\end{equation}
We can now use (\ref{eq:transdec}) to convert this into the affine decomposition in terms of Lemma~\ref{lem:decbdyas}, yielding,
\begin{equation}
 (\phi,T\phi)^{\bar{\mathrm{R}}}=\frac{1}{2}(\phi+T\phi)+(J\eta^{\mathrm{I}},-J\eta^{\mathrm{I}}), \quad
 (\phi,T\phi)^{\bar{\mathrm{I}}}=-\frac{1}{2}J(\phi-T\phi)-\eta^{\mathrm{I}} .
\end{equation}
In particular, setting $(\phi,T\phi)^{\bar{\mathrm{I}}}=0$ we find,
\begin{equation}
 T\phi=\phi-2J\eta^{\mathrm{I}} .
\label{eq:taf}
\end{equation}
Remembering the distinctions between the spaces $L_{\Sigma_1}$, $L_{\Sigma_2}$ etc.\ this takes the form,
\begin{equation}
 T\phi=\tilde{T}\phi-2 J_{\Sigma_2}\eta^{\mathrm{I}}_2 ,
\end{equation}
where $\phi\in L_{\Sigma_1}$, $T\phi\in L_{\Sigma_2}$ and $\eta^{\mathrm{I}}_2\defeq r_2(\eta^{\mathrm{I}})$.

Proposition~\ref{prop:aevol} now implies that the affine theory determined by $S_M^\mu$ leads to a unitary map $U:\cH_{\Sigma_1}\to\cH_{\Sigma_2}$ encoding the evolution from states on hypersurface $\Sigma_1$ to states on hypersurface $\Sigma_2$. For affine wave functions the explicit form of $U$ is given by equation (\ref{eq:unitindev}). Using also the identity (\ref{eq:aactsol}) this takes the form,
\begin{equation}
(U\psi)(\phi)=\exp\left(\frac{\im}{2} C_M^\mu\left(T^{-1}\phi,\phi\right)-\frac{\im}{2}[T^{-1}\phi,T^{-1}\phi]_{\Sigma_1}+\frac{\im}{2}[\phi,\phi]_{\Sigma_2}\right)\psi(T^{-1}\phi) .
\end{equation}
Since the boundary theory is linear it is natural to use the wave functions adapted to the linear setting instead. Using the relation (\ref{eq:wfbase}) between the settings yields for those wave functions,
\begin{equation}
(U\psi)(\phi)=\exp\left(\frac{\im}{2} C_M^\mu\left(T^{-1}\phi,\phi\right)\right)\psi(T^{-1}\phi) .
\end{equation}
The evolution of coherent states is given by equation (\ref{eq:evolcoh}). For the coherent states adapted to the linear setting, using (\ref{eq:relcohst}), this translates to,
\begin{equation}
U K_\xi = \exp\left(\frac{\im}{2} C_M^\mu\left(\xi,T\xi\right)+\frac{1}{4}g_{\Sigma_1}(\xi,\xi)-\frac{1}{4}g_{\Sigma_2}(T\xi,T\xi)\right) K_{T\xi} .
\end{equation}
The real part of the argument of the exponential can be seen to just compensate the different normalizations of the ``initial'' and ``final'' coherent states. Using normalized coherent states
\begin{equation}
\tilde{K}_\xi\defeq \exp\left(-\frac{1}{4}g_{\Sigma_1}(\xi,\xi)\right) K_\xi
\end{equation}
instead this yields,
\begin{equation}
U \tilde{K}_\xi = \exp\left(\frac{\im}{2} C_M^\mu\left(\xi,T\xi\right)\right) \tilde{K}_{T\xi} .
\end{equation}

In summary, the linear modification $C_M^\mu$ of the action $S_M$ causes a (normalized) coherent state to evolve to a (normalized) coherent state associated to a classical solution that is shifted by $-2 J\eta^{\mathrm{I}}$ compared to the original one, recall (\ref{eq:taf}). In particular, the vacuum state $\tilde{K}_0$ is shifted to the non-trivial coherent state $\tilde{K}_{-2 J \eta^{\mathrm{I}}}$, which is thus a superposition of states with all possible particle numbers. If $C_M^\mu$ represents a source term as in Section~\ref{sec:musrc} this is precisely the well known particle creation from the vacuum by a classical source. However, our analysis shows that this phenomenon generalizes far beyond theories with metric or causal background structures.

%% file: outlook.tex
\section{Discussion and Outlook}
\label{sec:outlook}

Besides providing a first generalization beyond linear field theory, there are more specific reasons for interest in affine field theory. One important reason was partially exploited in Section~\ref{sec:linsrc}. The addition of a linear term to a quadratic action (encoding a linear theory) yields an affine theory. If this linear term is taken to be a source term as in Section~\ref{sec:musrc}, this allows to set up a perturbation theory around the linear theory in the spirit of formula (\ref{eq:pertampl}). In Minkowski spacetime with a region enclosed by an initial equal-time hypersurface at $t_{\mathrm{initial}}$ and a final hypersurface at $t_{\mathrm{final}}$, this leads (together with other ingredients such as renormalization) in the asymptotic limit $t_{\mathrm{initial}}\to-\infty$ and $t_{\mathrm{final}}\to \infty$ just to the usual S-matrix. However, we have arrived here at this perturbative approach without using certain key ingredients that are normally part of the formalism of quantum field theory. In particular, the classical theory as encoded in the axioms of Section~\ref{sec:classdata} is not required to possess a metric or even causal background structure. We are thus in a position to set up and physically interpret a perturbation theory for field theories that possess no metric background. This suggests in particular the application to approaches to quantum gravity that are perturbative, but where the theory perturbed around is topological or at least not metrical, see e.g.\ \cite{FrSt:qgtopobs}.

Another reason for the specific interest in affine theories is that spaces of connections, which play a key role in gauge theories, are naturally affine spaces. That does not mean that the quantization scheme as presented in this paper is directly applicable to such theories. Indeed, in gauge theories the symplectic structure obtained along the lines of (\ref{eq:sympl}) is usually degenerate and one has to quotient by gauge transformations to obtain a non-degenerate form. In the present context one would have to take care, moreover, to perform this quotienting coherently for all the spaces of connections associated to the different hypersurfaces and regions. While this would allow the quantization of abelian gauge theories, the non-abelian case has further complications. In that case the symplectic structure is no longer invariant under translations in the affine space of connections. This case is thus quite beyond the quantization scheme discussed in the present paper. We note, however, that using a different quantization scheme the solvable case of non-abelian Yang-Mills theory in 2 dimensions has been shown to nicely incorporate into the GBF \cite{Oe:2dqym}.

The quantization prescription for amplitudes was motivated in Section~\ref{sec:feynquant} through the Feynman path integral. As mentioned there the form in which this motivation is presented is not fully satisfactory since the Feynman path integral is adapted to the Schrödinger rather than to the holomorphic representation. It thus becomes necessary to take a closer look at the Schrödinger representation and its relation to the holomorphic representation. It is indeed possible to establish this relation for affine field theory on a rigorous level.
This will be presented elsewhere.